\documentclass[12pt]{article}

\title{\Large\bf Non-commutative Calculus and Discrete Physics}

\author{Louis H. Kauffman \\
  Department of Mathematics, Statistics and Computer Science \\
  University of Illinois at Chicago \\
  851 South Morgan Street\\
  Chicago, IL, 60607-7045}

\begin{document}
 \maketitle
  
 \thispagestyle{empty}

 \section{Introduction} 
This paper is an expanded version of
\cite{NonCom} and \cite{Aspects} where there is presented an introduction to a point of view
for discrete foundations of physics. In taking a discrete stance, we find
that the initial expression of physical observation naturally occurs in a
context of non-commutative algebra. In
this way a formalism similar to quantum mechanics occurs first, but not
necessarily with the usual interpretations. By following this line
we show how the outlines of the well-known
forms of physical theory arise first in non-commutative form. The exact
relation of commutative and non-commutative theories raises a host of
problems.
\bigbreak

The starting point for this investigation is the representation of calculus in a 
non-commutative framework. In such a framework derivatives are represented by 
commutators, or more generally by products that satisfy the Jacobi identity and the Leibniz
rule. If we take commutators $[A,B] = AB -BA$ in an abstract algebra and define 
$DA = [A,J]$ for a fixed element $J$, then $D$ acts like a derivative in the sense that
$D(AB) = D(A)B + AD(B)$ (the Leibniz rule). As soon as we have calculus in such a framework,
concepts of geometry are immediately available. For example, if we have two
derivatives $\nabla_{J}A = [A,J]$ and $\nabla_{K}A =[A,K]$, then we can consider the commutator of these
derivatives $[\nabla_{J}, \nabla_{K}]A = \nabla_{J}\nabla_{K}A - \nabla_{K}\nabla_{J}A = [[J,K],A].$
The non-commutation of derivations corresponds to curvature in geometry, and indeed we shall see that the 
earliest emergence of curvature in this context is the formal analog of the curvature of a gauge connection!
\bigbreak

For multivariable calculus we need variables $X_1, X_2, \cdots, X_n$ and elements $P_1, P_2, ..., P_n$
such that $\partial_{i}A = \partial A/\partial X_{i} = [A, P_{i}].$ For a simplest representation  
we shall assume the the $X_i$ commute with one another, and that the $P_j$
commute with one another. Since we want $\partial_{i}X_j = \delta_{ij}$ (the Kronecker delta $\delta_{ij}$
is equal to one if $i$ and $j$ are equal and is zero otherwise), we must have the commutator equation
$[X_i,P_j] = \delta_{ij}.$ Thus multivariable calculus in this non-commutative representation demands
the commutation relations
$$[X_i,X_j] = 0$$
$$[P_i,P_j] = 0$$
$$[X_i,P_j] = \delta_{ij}$$ These equations are the ``flat background" for our non-commutative calculus.
The reader will note that this flat background has the same pattern of commutation relations as a
bare form of quantum mechanics when the $X$ variables are interpreted as position and the $P$ variables are
interpreted as momenta. In a certain sense this means that our considerations start in the quantum domain.
Note that flat is a correct adjective, since the derivatives $\partial_i$ all commute with one another.
\bigbreak

Let $A_i$ be a collection of elements of this algebra. Define ``covariant derivatives"
with $\Lambda_i = P_i - A_i$ by the formula 
$$\nabla_{i}Z = [Z, \Lambda_i] = \partial_{i}Z - [Z, A_i].$$ Computing the curvature, one finds
$$[\nabla_i,\nabla_j]Z = [[\Lambda_i,\Lambda_j], Z]$$ and
$$[\Lambda_i,\Lambda_j] = \partial_{i}A_j - \partial_{j}A_i + [A_i,A_j].$$ The reader will recognize this 
last expression as the formula for the curvature of a gauge connection.  
\bigbreak

\noindent In interfacing this formalism with physics we adopt the coupling equation 
$$dX_{i}/dt = \Lambda_{i} = P_{i} - A_{i}.$$ The reader will recognise this as the minimal coupling postulate
in the context of Poisson brackets. Here we take it in the context of commutators or Poisson brackets, or a more
general product satisfying the Jacobi identity and the Leibniz rule as described above. One retrieves the 
physics of a gauge field in this formalism. This is the essence of the pattern behind the 
Feynman-Dyson derivation of electromagnetism from commutation relations \cite{KN:QEM,Twist}, and its import
is more general. Because the brackets can be interpreted as commutators or as Poisson brackets with
special structure, the 
formalism can be seen in a multiplicity of contexts. Deeper relationships with curvature and metric
are related to this shifting of contexts as are relationships with quantum mechanics where the quantum formalism
is obtained by the Dirac prescription of replacing Poisson brackets by commutators. We will discuss these issues
in Section 5 of this paper. The organization of the paper is as follows.
\bigbreak

Section 2 of this paper we discuss  the properties of the
non-commutative discrete calculus that underlies our work. Here we begin with 
the consideration of a temporal operator $J$ with the property that $YJ = JY'$
for a ``time series" $X,X',X'',\cdots .$ Thus $XJ = JX', X'J = JX'', \cdots .$
This formalism for time series gives rise to the time derivative
$DA = [A,J] = AJ - JA = JA' - JA = J(A' - A),$ a commutator representing a discrete 
derivative. Note that $DA$ satisfies the Leibniz rule, a privilege not shared by the
usual commutative discrete derivative. This section discusses the discrete ordered calculus
(DOC) that arises from this idea and applies these ideas to a number of situations. In particular,
we consider the one variable case of the commutator equation $[X,DX]=Jk$ and show that it leads to 
a Brownian walk, and that if we take the size of the time step into account, then the diffusion constant for 
a Brownian process arises naturally as $k/2.$ We compare this with the usual derivation of the diffusion
constant and the diffusion differential equation. We then compare this situation with the one dimensional 
Schr\"{o}dinger equation, modeling it in relation to a diffusion process with complex amplitudes. In this
viewpoint one sees that the step length of the diffusion process is the Compton wavelength associated with 
the mass for the particle, and the time is the Compton time. For the Planck mass this gives a step equal to the 
Planck length and a time interval equal to the Planck time. We speculate on the relationship of this result to 
joint work with Pierre Noyes and others \cite{micro}. We consider other time series that can be regarded as solutions 
to this Heisenberg relation, the problem of using more variables and a model that is related to a discrete 
version of the Feynman-Dyson derivation of electromagnetic formalism.
\bigbreak

Section 3 examines the consequences for a particle whose position -
momentum commutator is equated to a metric field. Here we see how the
Levi-Civita connection (and implicitly differential geometric structure)
comes naturally from the non-commutative calculus. This is a very general result 
and in section 4 we discuss it in a more axiomatic context as described in this introduction.
This section discusses the intimate relationship between that Levi-Civita connection and the Jacobi 
and Leibniz identities that is revealed by our non-commutative calculus.
In section 5 our stance leads to an inversion of the usual
Dirac maxim ``replace Poisson brackets with commutators". If we replace
commutators with Poisson brackets that obey a Leibniz rule satisfied by
the commutators, then the dynamical variables will obey Hamilton's
equations. Thus we can take Hamilton's equations as a classicization of
our theory. Among other things, this point of view explains the appearance of the 
Levi-Civita connection in the abstract formalism. Interpreting with Poisson brackets,
we obtain a new proof (via Jacobi identity) of the classical result that a Newtonian particle
moving in generalized coordinates according to Lagrange's equations describes a geodesic in the
Levi-Civita connection.  Section 6 discusses the relationship of the discrete
ordered calculus with $q$-deformations and quantum groups. We show that
in a quantum group with a special grouplike element representing the
square of the antipode, there is a representation of the discrete ordered
calculus. In this calculus on a quantum group the square of the antipode
can represent one tick of the clock.  Then follows section 7 on networks and
discrete spacetime. This section is an exposition of ideas related to
spin networks and topological quantum field theory.  As an early example
we discuss the discretization of the Dirac equation in $1+1$ dimensional
spacetime. It is our speculation that the approaches to discrete physics
inherent in discrete calculus and in topological field theory are deeply
interrelated.  At the end of this section we outline this relationship in
the case of a model for quantum gravity due to Louis Crane. Section 8 is an appendix on 
the iterant approach to matrix algebra. We include this appendix to show how one can conceptualize
matrix algebra from point of view of the discrete. Section 9 is a philosophical appendix discussing the
nature of foundations in mathematics and in physics.
\vspace{3mm}

\noindent {\bf Remark.} The following references in relation to non-commutative calculus are useful in 
comparing with our approach \cite{Connes, Dimakis, Forgy, MH}. Much of the present work is the fruit of a long
series of discussions with Pierre Noyes, and we will be preparing collaborative papers on it. The present paper
is a summary for the proceedings of the ANPA Conference held in Cambridge, England in the summer of 2002.
I particularly thank Eddie Oshins for pointing out the relevance of minimal coupling. The
paper \cite{Mont} also works with minimal coupling for the Feynman-Dyson derivation. The first remark about
the minimal coupling occurs in the original paper by Dyson \cite{Dyson}, in the context of Poisson brackets.
The paper \cite{Hughes} is worth reading as a companion to Dyson. In the present paper we generalize the minimal
coupling to contexts including both commutators and Poisson brackets. The reader can see the full generality
of our approach by first reading this introduction and then going 
directly to sections 4 and 5. It is the purpose of this paper to indicate how non-commutative calculus
can be used in foundations.
\bigbreak 

\noindent {\bf Acknowledgement.} Most of this effort was sponsored by the Defense
Advanced Research Projects Agency (DARPA) and Air Force Research Laboratory, Air
Force Materiel Command, USAF, under agreement F30602-01-2-05022. Some of this
effort was also sponsored by the National Institute for Standards and Technology
(NIST). The U.S. Government is authorized to reproduce and distribute reprints
for Government purposes notwithstanding any copyright annotations thereon. The
views and conclusions contained herein are those of the author and should not be
interpreted as necessarily representing the official policies or endorsements,
either expressed or implied, of the Defense Advanced Research Projects Agency,
the Air Force Research Laboratory, or the U.S. Government. (Copyright 2003.)  
It gives the author pleasure to thank Pierre Noyes, Clive Kilmister, Ted Bastin, Tony Deakin,
Eddie Oshins, Basil Hiley, Keith Bowden,
Arleta Giffor, Ashok Gangadean, Lynnclaire Dennis, Louis Licht and Sam Lomonaco for many conversations during the course of this
work, and the Theory Group of the Stanford Linear
Accelerator Laboratory for hospitality during the preparation of parts of 
the present paper.
 \bigbreak

\section{Discrete Ordered Calculus}

In this section we recall the construction of an ordered version of the
calculus of finite differences $DOC$ \cite{KN:QEM}, \cite{NonCom}.  In
this calculus the Leibniz rule is satisfied, and so the calculus can be
used in a variety of applications. 
\vspace{3mm} 

In the abstract framework of this calculus, there are variables $X$, each
of which connotes a time series 
$$X,X',X'',....$$ Discrete unit time steps are indicated by the primes
appended to the $X$. A general point in the time series at time $t$ will
be denoted by $X^{t}$.  By convention let the time step between
successive points in the series be equal to 1 :  $$\Delta t = 1.$$ Then 
we can define the velocity at time $t$ by the formula:
$$v(t) = X^{t+1} - X^{t}.$$ More generally, if $X$ denotes position at a
given time, then $X'-X$ denotes the velocity {\em at that time}, where
the phrase ``at that time" must involve the next time as well. In a
discrete context there is no notion of instantaneous velocity.
\vspace{3mm} 

Measure position, and you find $X$. Then measure velocity, and you get
$X' - X$. Now measure position, and you get $X'$ because the time has
shifted to the next time in order to allow the velocity measurement. In
order to measure velocity the position is necessarily shifted to its
value at the next time step. In this sense, {\em position and velocity
measurements cannot commute in a discrete framework.} This is the key physical idea
that motivates our constructions. It was this idea, told to the author by Pierre Noyes,
that led to our papers and particularly to \cite{KN:QEM}.
\vspace{3mm}

The simplest interpretation of the variable $X$ is that the time series
values are numerical values, commuting with one another and with any
operators that might be present in the associated mathematics or physics.
In fact, we will often deal with situations where the $X$ and the
elements of the time series are in fact operators, not necessarily
commuting with one another.  At the very least we will construct an
algebra that mirrors the discrete non-commutativity of the operations of
position and velocity measurement.
\vspace{3mm}

Our project is to take this basic noncommutativity at face value and
follow out its consequences. To this end we will formulate a calculus of
finite differences that takes the order of observations into account.
This formalization is explained below.  
\vspace{3mm} 

To see most clearly the non-commutativity that is at the base of our considerations,
let $J$ denote the operation of shifting time by one increment. Thus we can envisage an 
algebra of operations that consists in commands like $JX$ (measure X, then tick the clock).
{\em Note that we will agree to take the sequence of operations from right to left.} Let
$|JX|$ denote the ``spatial evaluation" of this sequence of operations, obtained in general
by performing all the instructions and then evaluating the spatial position.  Thus
$$|JX| = X$$ while \noindent $$|XJ| = X'$$ \noindent since when the clock ticks, the
position shifts to the position at the next time. We see therefore, that $XJ \ne JX.$ This is
the first instance of non-commutativity in the physics of discrete space and time. From the
point of view of spatial evaluation it is most convenient to declare the equation
$$XJ = JX'$$ \noindent since these two expressions have identical spatial evaluations.
\bigbreak

We can then define the DOC derivative by the equation $$DX = [X,J] = XJ -JX = JX'-JX =
J(X'-X) = JdX$$ \noindent where $dX$ denotes the classical discrete derivative with unit
time step. The key point about the DOC derivative is that it is a commutator, and consequently
satisfies the Leibniz rule $$D(XY) = D(X)Y + XD(Y).$$ This makes it possible to do discrete
calculus in a way that is formally similar to classical calculus. We will repeat this 
structure more slowly now, first recalling the properties of classical discrete derivatives.
\bigbreak

We begin by recalling the usual derivative in the calculus of finite
differences, generalised to a (possibly) non-commutative context.
\vspace{3mm} 

\noindent {\bf Definition.} Let $$dX = X'-X$$ define the finite
difference derivative of a variable $X$ whose successive values in
discrete time are $$X,X',X'',....$$  This $dX$ is a classical derivative
in the calculus of finite differences. It is still defined even if the
quantities elements of the time series are in a non-commutative algebra. 
We shall assume that the values of the time series are in a possibly
non-commutative ring $R$ with unit. (Thus the values could be real
numbers, complex numbers, matrices, linear operators on a Hilbert space,
or elements of an appropriate abstract algebra.) This means that for
every element $A$ of the ring $R$ there is a well-defined successor
element $A'$, the next term in the time series. It is convenient to
assume that the ring itself has this temporal structure. In practice, one
is concerned with a particular time series and not the structure of the
entire ring.  Moreover, we shall assume that the next-time operator
distributes over both addition and multiplication in the sense that 
$$(A+B)' = A' + B'$$  and $$(AB)' = A'B'.$$   An element $c$  of the ring
$R$ is said to be a $constant$ if $c' = c.$ 
\vspace{3mm}

\noindent {\bf Lemma 1.} $$d(XY)= X'd(Y) + d(X)Y.$$
\vspace{3mm} 

\noindent {\bf Proof.} $$d(XY) = X'Y'-XY$$
$$=X'Y'-X'Y+X'Y-XY$$
$$=X'(Y'-Y) +(X'-X)Y$$
$$=X'd(Y) + d(X)Y.$$
\vspace{3mm}

This formula is {\em different} from the usual formula in Newtonian
calculus by the time shift of $X$ to $X'$ in the first term. We now
correct this discrepancy in the calculus of finite differences by taking
a {\em new} derivative $D$ as an {\em instruction to shift the time to
the left of the operator $D$.} That is, we take $XD(Y)$ quite literally
as an instruction to {\em first find $dY$ and then find the value of
$X.$} In order to find $dY$ the clock must advance one notch. Therefore
$X$ has advanced to $X'$ and we have that the evaluation of $XD(Y)$ is
$$X'(Y'-Y).$$
\vspace{3mm} 

In order to keep track of this non-commutative time-shifting, we will
write $$DX= J(X'-X)$$ where the element $J$ is a special time-shift
operator satisfying $$ZJ = JZ'$$ for any $Z$ in the ring $R$. The
time-shifter, $J$, acts to automatically evaluate expressions in the
resulting non-commutative calculus of finite differences. We call this
calculus $DOC$ (for discrete ordered calculus). Note that $J$ formalizes
the operational ordering inherent in our initial discussion of velocity
and position measurements. An operator containing $J$  causes a time
shift in the variables or operators to the left of $J$ in the sequence
order.
\vspace{3mm}

Formally, we extend the ring of values $R$ (see the definition of $d$
above) by adding a new symbol $J$ with the property that $AJ = JA'$ for
every $A$ in $R$. It is assumed that the extended ring $R$ is associative
and satisfies the distributive law so that $J(A+B) = JA + JB$ and $J(AB)
= (JA)B$ for all $A$ and $B$ in the ring. We also assume that $J$ itself
is a constant in the sense that $J' = J$.
\vspace{3mm}

The key result in $DOC$  is the following adjusted difference formula: 
\vspace{3mm}

\noindent {\bf Lemma 2.} $$D(XY) = XD(Y) + D(Y)X.$$ 
\vspace{3mm}

\noindent {\bf Proof.} $$D(XY)$$
$$ = J(X'Y'-XY)$$
$$= J(X'Y'-X'Y +X'Y-XY)$$
$$=J(X'(Y'-Y) + (X'-X)Y$$
$$=JX'(Y'-Y) + J(X'-X)Y$$
$$=XJ(Y'-Y) + J(X'-X)Y$$
$$=XD(Y) + D(X)Y.$$
\vspace{3mm} 

The upshot is that $DOC$ behaves formally like infinitesimal calculus and
can be used as a calculus in this version of discrete physics. In
\cite{KN:QEM} Pierre Noyes and the author use this foundation to build a
derivation of  a non-commutative version of electromagnetism.  Another
version of this derivation can be found in \cite{Twist}.  In both cases
the derivation is a translation to this context of the well-known
Feynman-Dyson derivation of electromagnetic formalism from commutation
relations of position and velocity. 
\vspace{3mm}

 Note that the  definition of the derivative in $DOC$  is actually a
commutator: $$DX = J(X'-X) = JX' - JX = XJ -JX = [X,J].$$  The operator
$J$ can be regarded as a discretised time-evolution operator in the
Heisenberg formulation of quantum mechanics. In fact we can write
formally that $$X' = J^{-1}XJ$$ since $JX' = XJ$ (assuming for this
interpretation that the operator $J$ is invertible). Putting the time
variable back into the equation, we get the evolution
$$X^{t+ \Delta t} = J^{-1}X^{t}J.$$  This aspect can be compared to the 
formalism of Alain Connes' theory of non-commutative geometry
\cite{Connes}.
\vspace{3mm}

In $DOC$,  $X$ and $DX$ have no reason to commute: $$[X,DX] = XJ(X'-X)
-J(X'-X)X = J(X'(X'-X) -(X'-X)X)$$ Hence $$[X,DX] = J(X'X'-2X'X + XX).$$

\noindent This is non-zero even in the case where $X$ and $X'$ commute with
one another. Consequently, we can consider physical laws in the form 
$$[X_{i}, DX_{j}] = g_{ij}$$ where $g_{ij}$ is a function that is
suitable to the given application. In \cite{KN:QEM} we show how the
formalism of electromagnetism arises when $g^{ij}$ is $\delta^{ij}$, the
Kronecker delta. In \cite{KN:DG} we will show how  the general case
corresponds to a ``particle" moving in a non-commutative gauge field
coupled with geodesic motion relative to the Levi-Civita connection
associated with the $g_{ij}.$ This result can be used to place the work
of Tanimura \cite{Tanimura} in a discrete context. 
\vspace{3mm}

It should be emphasized that all physics that we derive in this way is
formulated in a context of non-commutative operators and variables. We do
not derive electromagnetism, but rather a non-commutative analog.  It is
not yet clear just what these non-commutative physical theories really
mean. Our initial idealisation of measurement is not the only model for
measurement that corresponds to actual observations. Certainly the idea
that we can measure time  in a way that has ``steps between the steps of
time" is an idealisation. It happens to be an idealisation that fits a
model of the universe as a cellular automaton. In a cellular automaton an
observation  is what an operator of the automaton might be able to do. It
is not necessarily what the ``inhabitants" of the automaton can perform.
Here is the crux of the matter. The inhabitants can have only limited
observations of the running of the automaton, due to the fact that they
themselves are processes running on the automaton. The theories we build
on the basis of $DOC$ can be theories {\em about} the structure of these
automata. They will eventually lead to theories of what can be observed
by the processes that run on such automata. It is possible that the well
known phenomena of quantum mechanics will arise naturally in such a
context. These points of view should be compared with \cite{Fkin}.
\vspace{3mm} 

\subsection{Brownian Walks and the Diffusion Equation}
To return to basics, consider the commutator equation in one space variable $X.$ 

$$[X, DX] = Jk$$

\noindent
 for a single variable $X.$   Written out, this equation becomes

$$Jk = [X, J(X'-X)] = XJ(X'-X) - J(X'-X)X = J(X'(X'-X) - (X'-X)X).$$

\noindent If $k$ and the elements of the time series $\{X,X',X'',...\}$
are all commuting scalars then this equation reduces to 

$$k = (X'-X)^{2}.$$

\noindent Thus $$X' = X \pm k^{1/2},$$  

\noindent a Brownian random walk, is a
solution to the simplest one-dimensional commutator equation.
\bigbreak

Now lets examine this Brownian walk more closely by quantifying the time step as well 
as the space step. We take $$\Delta t = \tau$$ \noindent so that 
$$DX = J(X'-X)/\tau$$ \noindent where it is assumed that $\tau$ is a scalar, commuting with 
all elements of the time series and commuting with the operator $J$ (that is, $\tau$ does
not change with time). Now examine once again the equation  $$[X, DX] = Jk.$$ 
\bigbreak

\noindent Let $|X'-X| = \Delta.$ Then, repeating the calculation, we find 
$$k = (X'-X)^{2}/\tau =\Delta^{2}/\tau.$$ \noindent Hence $$\Delta^{2}/\tau = k.$$
\noindent This tells us that if $k$ is to be constant then there must be a constant
relationship between the square of the space interval for the Brownian walk and the size of 
the time interval. The remarkable point here is that it is just this constant relationship
that is required for a Brownian process to be described by the diffusion equation
$$\partial P(x,t)/\partial t  = C \partial^{2}P(x,t)/\partial x^{2}$$
\noindent where the diffusion constant $C$ is given by the formula 
$$C = \Delta^{2}/2 \tau = k/2.$$  \noindent The diffusion constant comes 
directly from our consideration involving the DOC commutator without any of the usual 
conceptual apparatus about approximating a differential equation.
\bigbreak

To make this comparison, lets recall how the diffusion equation usually arises in discussing
Brownian motion. We are given a Brownian process where $$x(t + \tau) = x(t) \pm \Delta$$ 
\noindent so that the time step is $\tau$ and the space step is of absolute value $\Delta.$
We regard the probability of left or right steps as equal, so that if $P(x,t)$ denotes the
probability that the Brownian particle is at point $x$ at time $t$ then
$$P(x, t+\tau) = P(x-\Delta,t)/2 + P(x+\Delta)/2.$$ \noindent From this equation for the probability
we can write a difference equation for the partial derivative of the probability with respect
to time: 
$$[(P(x, t + \tau) - P(x,t))/\tau] = (h^{2}/2\tau)[(P(x-\Delta,t) - 2P(x,t) + P(x+\Delta))/\Delta^{2}] $$ 
\noindent The expression in brackets on the right hand side is a discrete approximation to
the second partial of $P(x,t)$ with respect to $x.$ Thus if the ratio $C = \Delta^{2}/2\tau$
remains constant as the space and time intervals approach zero, then this equation goes in 
the limit to the diffusion equation 
$$\partial P(x,t)/\partial t  = C \partial^{2}P(x,t)/\partial x^{2}.$$
\bigbreak

It is most curious how the diffusion constant comes up in these two contexts. Lets try to
think about the comparison between the non-commutative observational starting point and 
the more standard differential approximation. In the non-commutative context we get $\Delta^{2}$
from the appearance of the square of the difference of $X'$ and $X$ in the calculation of 
the commutator of $X$ and $DX.$ In the differential approximation, we get the $\Delta^{2}$ from
the approximation of the second derivative of the probability $P(x,t)$ with respect to $x.$
The concept of probability does not appear in the non-commutative context. Clearly
this subject needs more thought.
\bigbreak

\subsection{Planck's Numbers, Schr\"{o}dinger's Equation and the Diffusion Equation}
First recall the Planck Numbers. $\hbar$ is Planck's constant divided by $2\pi.$
$c$ is the speed of light.
$G$ is Newton's gravitational constant. The Planck length will be denoted by $L$, the Planck
time by $T$ and the Planck mass by $M.$ Their formulas are
$$M = \sqrt{\hbar c/G}$$
$$L = \hbar/Mc$$
$$T = \hbar/Mc^{2}.$$
\noindent These amounts of mass, length and time have just these dimensions and are constructed
from the values of fundamental physical constants. They have roles in physics that point
to deeper reasons than the formal for introducing them. Here we shall see how they are related
to the Schr\"{o}dinger equation.
\bigbreak

Recall that Schr\"{o}dinger's equation can be regarded as the diffusion equation with an 
imaginary diffusion constant. Recall how this works. The Schr\"{o}dinger equation is
$$i\hbar \partial \psi/\partial t = H\psi$$ \noindent where the Hamiltonian $H$ is given by
the equation $H = p^{2}/2m + V$ where $V(x,t)$ is the potential energy and 
$p = \hbar/i \partial/\partial x$ is the momentum operator. With this we have
$p^{2}/2m =  (-\hbar^{2}/2m) \partial^{2}/\partial x^{2}.$ 
Thus with $V(x,t) = 0$, the equation becomes 
$i\hbar \partial \psi/\partial t = (-\hbar^{2}/2m) \partial^{2} \psi/\partial x^{2}$
which simplifies to
$$\partial \psi/\partial t = (i\hbar/2m) \partial^{2} \psi/\partial x^{2}.$$
\noindent Thus we have arrived at the form of the diffusion equation with an imaginary 
constant, and it is possible to make the identification with the diffusion equation by 
setting $$\hbar/m = \Delta^{2}/\tau$$ \noindent where $\Delta$ denotes a space interval, and
$\tau$ denotes a time interval as explained in the last section about the Brownian walk.
With this we can ask what space interval and time interval will satisfy this relationship with 
a mass and Planck's constant? {\em Remarkably, the answer is that this equation is satisfied when
$m$ is the Planck mass, $\Delta$ is the Planck length and $\tau$ is the Planck time!!}
For note that $$L^{2}/T = (\hbar/Mc)^{2}/(\hbar/Mc^{2}) = \hbar/M.$$
\bigbreak

I now quote an email comment of Pierre Noyes:
``With regard to your DOC derivation of the diffusion equation,
and with an imaginary diffusion coefficient, the Schr\"{o}dinger equation, note that
the relation $\hbar/m=L^2/T$ is satisfied for {\em any} mass $m$ provided we take $L$ = Compton
wavelength = $\hbar/mc$ and $T$ =Compton time= $\hbar/mc^2$ --- which is simply the time of
a step length of this length taken at the velocity of light. I have a vague idea
that I heard of this relation when I was a graduate student. In any case I am
sure Feynman had it in mind when he used a random walk on the light cone to
derive the $1+1$ Dirac equation, and counted steps using $i$! So, in a sense, your
DOC derivation of the diffusion equation does connect the Maxwell equations
derivation via DOC, to the Dirac equation derivation --- which in a vague sense
was what I hoped we would be able to do this spring (2002).
Of course this general result applies in particular to the Planck mass, which was
your first observation. It is intriguing that if the mass scale is $m$ [the Planck mass], then we can
use either the Compton wavelength or the Schwarzchild radius at that mass scale
as the step length in DOC. This reinforces my conviction (expressed long ago)
that elementary particles are small black holes."
\bigbreak

The last part of Noyes' remark about the Schwarzchild radius refers to our work \cite{micro}
explaining Ed Jones' microcosmology.  Jones observed that if, for a particle of mass $m$
we set the Schwarzchild radius ($R_{S} = 2mG/c^{2}$) equal to the Compton radius
($R_{C} = \hbar/2mc$), then the resulting mass $m$ is equal 
to one half the Planck mass!
$$R_{S} = R_{C}$$
$$2mG/c^{2} =  \hbar/2mc$$
$$m =(1/2) \sqrt{\hbar c/G} = M/2$$
\noindent This is highly suggestive of limiting conditions on matter (``Plancktonic matter")
prior to the Big Bang and leads in this way to specific cosmological predictions. It also
gives an intriguing physical meaning to the Planck mass.
\bigbreak

What does all this say about the nature of the Schr\"{o}dinger equation itself?
Interpreting it as a diffusion equation with imaginary constant suggests comparing with the
DOC equation $$[X,DX]=JiC$$ for a real constant $C$. This equation implicates a Brownian process
where $X'= X \pm Z$ where $Z^{2}/ \tau = iC.$ We can take $Z = \sqrt{i}L$ where $L$ is a real
step-length. This gives a Brownian walk in the complex plane with the correct DOC diffusion 
constant. However, the relationship of this walk with the Schr\"{o}dinger equation is less 
clear because the $\psi$ in that equation is not the probability for the Brownian process.
To see a closer relationship we will take a different tack.
\bigbreak

Consider a discrete function $\psi(x,t)$ defined (recursively) by the following equation
$$\psi(x, t+\tau) = (i/2)\psi(x - \Delta,t) + (1-i)\psi(x,t) + (i/2)\psi(x + \Delta,t)$$
\noindent In other words, we are thinking here of a random ``quantum walk" where the amplitude
for stepping right or stepping left is proportional to $i$ while the amplitude for not moving
at all is proportional to $(1-i).$ It is then easy to see that $\psi$ is a discretization of 
$$\partial \psi/\partial t = (i\Delta^{2}/2\tau)\partial^{2} \psi/\partial x^{2}.$$
\noindent Just note that $\psi$ satisfies the difference equation
$$(\psi(x,t+\tau) - \psi(x,t))/\tau =
(i\Delta^{2}/2\tau)(\psi(x - \Delta,t) -2\psi(x,t) + \psi(x + \Delta,t))/\Delta^{2}$$
\noindent This gives a direct interpretation of the solution to the 
Schr\"{o}dinger equation as a limit of a sum
over generalized Brownian paths with complex amplitudes. We can then reinterpret this in 
DOC terms by the equation $[X,DX] = J(\Delta^{2}/\tau)$ or $[X,DX] = 0$, each of these
contingincies happening probabilistically. It remains to be seen whether there is further 
insight to be gained into the Schr\"{o}dinger equation via this combination of the DOC approach
and the stochastic approach.  
\bigbreak

\subsection{DOC Chaos}
Along with the simple Brownian motion solution to the one dimensional commutator equation, there is a 
heirarchy of time series that solve this equation, with periodic and chaotic behaviour. These solutions
can be obtained by taking 
$$X= J^{n}Y$$ where Y is a numerical scalar, and taking the commutator equation to be
$$[X, DX] = J^{2n+1}k$$ where $k$ is a scalar. Expanding this equation, we find
$$XJ(X'-X) - J(X'-X)X = J^{2n+1}k$$
$$J^{n}YJ(J^{n}Y' - J^{n}Y) - J(J^{n}Y' - J^{n}Y)J^{n}Y = J^{2n+1}k$$ 
$$J^{2n+1}Y^{n+1}(Y' - Y) - J^{2n+1}(Y^{n+1} - Y^{n})Y = J^{2n+1}k$$ 
$$Y^{n+1}(Y' - Y) - (Y^{n+1} - Y^{n})Y = k$$ 
$$Y^{n+1}(Y' - 2Y) = k - Y^{n}Y $$ 
$$Y^{n+1} = (k - Y^{n}Y)/(Y'-2Y).$$ This last equation expresses the time series recursively where $Y$ refers
to the value of the series that is $n$ time steps back from $Y^{n}.$ The first case of this recursion is
$$Y'' = (k - Y'Y)/(Y'-2Y).$$ Next case is
$$Y''' = (k - Y''Y)/(Y'-2Y).$$ These recursions depend critically on the value of the parameter $k.$ In the first
case one sees periodic oscillations that (for appropriate values of $k$) destabilize and blow up, alternating 
between an unbounded phase and a bounded semi-periodic phase. We will investigate these time series in a 
separate paper.
\bigbreak

\subsection{More Variables}
In the Feynman-Dyson derivation of electromagnetic
formalism from commutation relations \cite{KN:QEM} one uses the relations 
$$[X_i, X_j] = 0$$
$$[X_i, DX_j] = k\delta_{ij}$$
\noindent where $k$ is a scalar. Here we shall use
$$[X_i, X_j] = 0$$
$$[X_i, DX_j] = Jk\delta_{ij}$$ \noindent as we did in analyzing the one-dimensional case.
This allows us to have scalar evolution of the time series, but changes some of the issues in the Feynman-Dyson
derivation. These are in fact handled by the more general formalism that we discuss in the next two sections.
Thus we shall aim in this section to see to what extent one can make simple models for this version of the 
Feynman-Dyson relations. Models of this sort will be another level of approximation to  discrete 
electromagnetism. 
\bigbreak 

Writing out the commutation relation $[X,DX]=Jk$, and not making
any assumption that $X'$ commutes with $X$, we find
$$J^{-1}[X,DX] = X'(X'-X) -(X'-X)X$$
$$= X'(X'-X) -X(X'-X) +X(X'-X) -(X'-X)X$$
$$= (X'-X)^2 + (XX'-X'X) = (X'-X)^2 + [X,X'].$$

\noindent Thus the commutation relation $[X,DX]=Jk$ becomes the equation
$$(X'-X)^2 + [X,X'] = k.$$

\noindent By a similar calculation, the equation $[X,DY]=0$ becomes the equation
$$(X'-X)(Y'-Y) + [X,Y']=0.$$

These equations are impossible to satisfy simultaneously for $k \ne 0$ 
if we assume that $X$ and $X'$ commute and that 
$X$ and $Y'$ commute and that $[Y,DY]=Jk$. For then we would need to solve:
$$(X'-X)^2 = k.$$
$$(Y'-Y)^2 = k.$$
$$(X'-X)(Y'-Y) =0.$$
\noindent with the first two equations implying that $(X-X')$ and $(Y-Y')$ are each
non-zero, and the third implying that their product is equal to zero.  
In other words, the equations below cannot be satisfied if the time series are composed of
commuting scalars. 
$$[X,DX]=Jk$$
$$[Y,DY]=Jk$$
$$[X,Y]=0$$ \noindent
In order to make such models we shall have to introduce non-commutativity into
the time series themselves. In a certain sense this is analoguous to the introduction of 
non-commutative algebra in the Dirac equation in $3+1$ dimensions, and to the introduction of 
non-commutative fields in gauge theory. 
\bigbreak

\noindent Here is an example of such a model.
\smallbreak
\noindent Return to the equations
$$(X'-X)^2 + [X,X'] = k.$$
$$(X'-X)(Y'-Y) + [X,Y']=0$$ \noindent expressing the behaviour for two distinct variables $X$ 
and $Y.$ If $[X,X'] = 0$, then we have $(X'-X)^2 = k$ so that 
$$X' = X \pm \sqrt{k}.$$  

\noindent In order for the second equation to be satisfied, we need that 
$$[X,Y'] = \pm k$$ \noindent where the ambiguity of sign is linked with the varying signs in the 
temporal behaviour of $X$ and $Y.$ We will make the sign more precise in a moment, but the radical 
part of this suggestion is that for two distinct spatial variables $X$ and $Y$, there will be a 
commutation relation between one and a time shift of the other.
\bigbreak

\noindent If the space variables are labeled $X_i$, then we can write 
$$X_{i}^{t+1} = X_{i}^{t} + \epsilon_{i}^{t} k$$ \noindent where $\epsilon_{i}^{n}$
is plus one or minus one. Thus each space variable performs a walk with the fixed step-length
$k.$  We shall write informally $$X_{i}' = X_{i} + \epsilon_{i} k$$ \noindent where it is 
understood that the epsilon without the superscript connotes the sign change that occurs in this
juncture of the process. We then demand the commutation relations
$$[X_{i}', X_{j}] = [X_{j}', X_{i}] = \epsilon_{i} \epsilon_{j} k.$$ \noindent Each $X_{i}$ is
a scalar in its own domain, but does not commute with the time shifts of the other directions.
We then can have the full set of commutation relations:
$$[X_{i}', X_{j}] = [X_{j}', X_{i}] = \epsilon_{i} \epsilon_{j} k.$$
$$[X_i, X_j] = 0$$
$$[X_i, DX_j] = Jk\delta_{ij}$$ \noindent so that the system will satisfy the assumptions 
supporting the Feynman-Dyson derivation. In this system,the elements of a given time series
$X_{i}, X_{i}',X_{i}'', \cdots$ commute with one another. The basic field element in the Feynman-Dyson
set up is the magnetic field $B$ defined by the (non-commutative) vector cross product
$$B =(1/k)  DX \times DX.$$ \noindent Here we have
$$DX_{i} = J(X_{i}' - X_{i}) = J \epsilon_{i} \sqrt{k}.$$ \noindent Thus $$B = J^{2} ~ \epsilon' \times \epsilon$$
\noindent where $\epsilon =(\epsilon_{1}, \epsilon_{2}, \epsilon_{3})$ (assuming three spatial coordinates) and 
$\epsilon'$ denotes this vector of signs at the next time step. In this way we see that we can think of each
spatial coordinate as providing a long temporal bit string and the three coordinates together give the field in
terms of the vector cross product of their temporal cross sections at neighboring instants. It is interesting to
compare this model with the color algebra in the following paper by Wene \cite{Wene}. 
\bigbreak

\subsection{Discrete Classical Electromagnetism}
It is of interest to compare these results with a direct discretization of classical electromagnetism.
Suppose that $X,X',X'',X''', \cdots$ is a time series of vectors in $R^{3}$ (where $R$ denotes the real numbers).
Let $dX = X' - X$ be the usual discrete derivative (with time step equal to one for convenience).
Let $A \bullet B$ denote the usual inner product of vectors in three dimesions.
Assume that there are fields $E$ and $B$ such that $$ d^{2}X = E + dX \times B$$ \noindent
(the Lorentz force law). Assume also that $E$ and $B$ are perpendicular to the velocity vector $dX$, and 
that $E$ is perpendicular to $B.$ Then we have
$$dX' \times dX = (dX' - dX) \times dX =
 (d^{2}X) \times (dX)$$
$$= E \times dX + (dX \times B) \times dX$$
$$= E \times dX - dX (B \bullet dX) + (dX \bullet dX)B.$$ \noindent
Since $E$ is perpendicular to $dX$ we know there is a $\lambda$ such that $E \times dX = \lambda B$
and we have $B \bullet dX = 0$ since $B$ is perpendicular to $dX.$ Therefore
$$dX' \times dX = \lambda B + ||dX||^{2} B$$ \noindent so that
$$B = dX' \times dX /(\lambda + ||dX||^{2}).$$ \noindent Up to the factor in the denominator,
this formula is in exactly the same pattern as the formula in our discrete model for DOC electromagnetism as
described in the previous subsection. To see this, note that the $B$ field in the DOC model is proportional to
$DX \times DX$ and that $DX = JdX$ so that $DX \times DX = JdX \times JdX = J^{2}dX' \times dX.$ Up to the time-shifting
algebra and a proportionality constant, the expressions are the same!
Clearly more work is needed in comparing classical discrete electromagnetism
with the results of a discrete analysis of the Feynman-Dyson derivation.
\bigbreak

\section{Gauge Fields and Differential Geometry}

Letting $X_{i}$ ($i=1,2,...,d$) denote a set of spatial variables
(non-commutative time series in the sense of our discrete ordered
calculus), we will look at a collection of basic assumptions about the
commutation of these variables and of their derivatives. It is natural
from the point of view of the discrete ordered calculus to have
$$[X_{i}, X_{j}] = 0$$  for all $i$ and $j$.  There are no other natural
commutations from the point of view of this calculus.
\vspace{3mm}

We shall define $g_{ij}$ by the equation
$$[X_{i}, \dot{X_{j}}] = g_{ij}.$$  Here $\dot{X_{j}}$ is shorthand
for$DX_{j}$ and 
$$[A,B] = AB -BA.$$
\vspace{3mm}

 Along with this commutator equation, we will assume that 
$$[X_{i}, X_{j}] = 0,$$
$$[X_{i}, g_{jk}] = 0$$ and 
$$[g_{rs}, g_{jk}] =0.$$

\noindent Here it is assumed that  $g_{ij}$ is non-degenerate in the
sense that there exists $g^{ij}$ so that $$g^{ij}g_{jk} =
\delta^{i}_{k}$$ and that 
$$g_{ij}g^{jk} = \delta_{i}^{k}.$$  Here we are using the Einstein
summation convention that implicitly assumes that we sum over repeated
indices in an expression. Symbol $\delta^{i}_{j}$ is a Kronecker delta,
equal to $1$ when $i$ equals $j$ and $0$ otherwise.   
\vspace{3mm}

The first result that is a direct consequence of these assumptions is the
symmetry of the ``metric" coefficients $g^{ij}.$  That is, we shall show
that 
$$g^{ij}=g^{ji}.$$
\vspace{3mm}

\noindent {\bf Lemma 3.} $g_{ij}=g_{ji}.$
\vspace{2mm}

\noindent {\bf Proof.}
$$g_{ij} - g_{ji}$$ 
$$= [X_{i}, \dot{X_{j}}]  - [X_{j}, \dot{X_{i}}] $$
$$= [X_{i}, \dot{X_{j}}]  + [\dot{X_{i}}, X_{j}] $$
$$= D[X_{i}, X_{j}]$$
$$ = 0.$$
\vspace{3mm}

For the purpose of doing calculus in this situation we define
$\dot{X^{i}}$ by the equation

$$\dot{X^{i}} = g^{ik} \dot{X_{k}}.$$

\noindent The operator $\dot{X^{i}}$ is simply the index shift of the
corresponding $\dot{X_{i}}.$  We do not define a corresponding $X^{i}.$
It is easy to check the equation

$$[X_{i}, \dot{X^{j}}] = \delta^{j}_{i}.$$

\noindent Consequently, we define the derivative of an operator $F$ with
respect to $X_{i}$ by the equation

$$\partial^{i}F = [F, \dot{X^{i}}]$$

\noindent and the corresponding lowered derivative by the formula

$$\partial_{i}F = [F, \dot{X_{i}}].$$

\noindent Note that we have

$$\partial_{i}X_{j} = g_{ij}.$$

\noindent We also define $$\hat{\partial_{i}}F = [X_{i}, F],$$ the derivative of $F$ with respect 
to the conjugate variable $\dot{X^{i}}.$
\bigbreak

\noindent With these partial derivatives in hand, we define $\dot{F}$ by
the formula

$$\dot{F} = \partial^{k}F\dot{X_{k}}.$$

\noindent If $F$ commutes with $g^{ij}$ then it is easy to see that 

$$\dot{F} = \partial_{k}F\dot{X^{k}}.$$

\noindent These formulas extend (implicitly) the definition of the time
series to entities other than the operators $X_{i}$  since $$\dot{F} = DF
= J(F' -F).$$
\vspace{3mm}

A stream of consequences then follows by differentiating both sides of
the equation 

$$g_{ij}= [X_{i}, \dot{X_{j}}].$$ 

\noindent Note that  

$$\dot{g_{ij}}= [\dot{X_{i}}, \dot{X_{j}}] + [X_{i}, D^{2}X_{j}]$$

\noindent by the Leibniz rule

$$D[A,B] = [DA.B] + [A,DB].$$

\noindent Note also that we can freely use the Jacobi identity

$$[A, [B,C]] + [C, [A,B]] + [B, [C,A]] = 0.$$

\vspace{3mm}

In particular, the Levi-Civita connection 

$$\Gamma_{ijk} =(1/2)(\partial_{i}g_{jk}
+\partial_{j}g_{ik}-\partial_{k}g_{ij})$$  

\noindent associated with the $g_{ij}$ comes up almost at once from the
differentiation process described above.  To see how this happens, view
the following calculation where 
$$\hat{\partial_{i}}\hat{\partial_{j}}F = [X_{i}, [ X_{j}, F]].$$
\vspace{3mm}

We apply the operator $\hat{\partial_{i}}\hat{\partial_{j}}$ to the second $DOC$ derivative of
$X_{k}.$
\vspace{3mm}

\noindent {\bf Lemma 4.}  $\Gamma_{ijk}= (1/2)\hat{\partial_{i}}\hat{\partial_{j}}D^{2}X_{k}$ 
\vspace{2mm}

\noindent {\bf Proof.} 

$$\hat{\partial_{i}}\hat{\partial_{j}}D^{2}X_{k} = [X_{i}, [ X_{j}, D^{2}X_{k}]]$$

$$= [X_{i}, \dot{g_{jk}} - [\dot{X_{j}}, \dot{X_{k}}]]$$

$$= [X_{i}, \dot{g_{jk}}] -  [X_{i}, [\dot{X_{j}}, \dot{X_{k}}]]$$

$$= [X_{i}, \dot{g_{jk}}] + [\dot{X_{k}}, [X_{i}, \dot{X_{j}}]]+
[\dot{X_{j}}, [ \dot{X_{k}}, X_{i}]$$

$$= [g_{jk}, \dot{X_i}] +  [\dot{X_{k}}, g_{ij}] + 
[\dot{X_{j}}, -g_{ik}]$$

$$= \partial_{i}g_{jk} -\partial_{k}g_{ij} + \partial_{j}g_{ik}$$

$$ = 2\Gamma_{kij}.$$

\noindent It is remarkable that the form of the Levi-Civita connection
comes up directly from this non-commutative calculus without any apriori
geometric interpretation. We shall discuss the context of this result in the next two sections of
the paper.
\vspace{3mm}

One finds that 
$$D^{2}X_{i} = G_{i} + g_{ir}g_{js}F^{rs}\dot{X^{j}}
+\Gamma_{ijk}\dot{X^{j}} \dot{X^{k}}$$

\noindent where 
$$F^{rs} = [\dot{X^{r}}, \dot{X^{s}}].$$

\noindent It follows from the Jacobi identity that 

$$F_{ij}=g_{ir}g_{js}F^{rs}$$

\noindent
 satisfies the equation 
$$\partial_{i}F_{jk} + \partial_{j}F_{ki} + \partial_{k}F_{ij} = 0,$$
identifying $F_{ij}$ as a non-commutative analog of a gauge field. $G_{i}$
is a non-commutative analog of a scalar field.   The details of these
calculations will be found in \cite{KN:DG}.
\vspace{3mm}

This description of the equations for a non-commutative particle in a
metric field illustrates the role of the background discrete time in this
theory. In terms of the background time the metric coefficients are not
constant. It is through this variation that the spacetime derivatives of
the theory are articulated.   The background is a process with its own
form of discrete time, but no spacetime structure as we know and observe
it. Our observation of spacetime structure appears as a rough
(commutative) approximation to the processes described as consequences of
the basic non-commutative equations of the discrete ordered calculus.  
\vspace{3mm}

\section{Curvature, Jacobi Identity and the Levi-Civita Connection}
In this section, we go back to basics and examine the context of calculus defined via 
commutators. We shall use a partially index-free notation. 
In this notation, we avoid nested subscripts by using different variable names and then using these 
names as subscripts to refer to the relevant variables. Thus we write $X$ and $Y$ instead of
$X_{i}$ and $X_{j}$, and we write $g_{XY}$ instead of $g_{ij}.$ It is assumed that the derivation
$DX$ has the form $DX = [X,J]$ for some $J.$
\bigbreak

The bracket $[A,B]$ is not assumed to be a commutator. It is assumed to satisfy
the Jacobi identity, bilinearity in each variable, and the Leibniz rule for all functions of the
form $\delta_{K}(A) = [A,K].$ That is we assume that 
$$\delta_{K}(AB) =  \delta_{K}(A)B + A\delta_{K}(B).$$
\bigbreak

Recall that in classical differential
geometry one has the notion of a covariant dervative, defined by taking a difference quotient
using parallel translation via a connection. Covariant derivatives in different directions do not 
necessarily commute. The commutator of covariant derivatives gives rise to the curvature tensor
in the form $$[\nabla_{i}, \nabla_{j}]X^{k} = R^{k}_{lij}X^{l}.$$ If derivatives do not commute
then we regard their commutator as expressing a curvature. In our non-commutative context this 
means that curvature arises {\em prior} to any notion of covariant derivatives since 
{\em even the
basic derivatives do not commute.}
\bigbreak 

We shall consider derivatives in the form $$\nabla_{X}(A) = [A, \Lambda_{X}].$$

\noindent Examine the following computation:
$$\nabla_{X}\nabla_{Y}F = [[F,\Lambda_{Y}],\Lambda_{X}] = -[[\Lambda_{X}, F],\Lambda_{Y}]
 - [[\Lambda_{Y},\Lambda_{X}],F]$$
$$= [[F, \Lambda_{X}],\Lambda_{Y}] + [[\Lambda_{X},\Lambda_{Y}],F]$$
$$= \nabla_{Y}\nabla_{X}F + [[\Lambda_{X},\Lambda_{Y}],F].$$
\noindent Thus
$$[\nabla_{X}, \nabla_{Y}]F= R_{XY}F$$
\noindent where $$R_{XY}F = [[\Lambda_{X},\Lambda_{Y}],F].$$
\noindent We can regard $R_{XY}$ as a curvature operator. 
\bigbreak

The analog in this context of flat space is abstract quantum mechanics! 
That is, we assume position variables (operators)
$X$, $Y$, $\cdots$ and momentum variables (operators) $P_X$, $P_Y$, $\cdots$ satisfying the 
equations below.
$$[X,Y]=0$$
$$[P_X,P_Y]=0$$
$$[X, P_Y] = \delta_{XY}$$
\noindent where $\delta_{XY}$ is equal to one if $X$ equals $Y$ and is zero otherwise.
We define $$\partial_{X}F = [F,P_X]$$ \noindent and $$\partial_{P_X}F = [X,F].$$ 
\noindent In the context of the above commutation relations, note that these derivatives
behave correctly in that $$\partial_{X}(Y) = \delta_{XY}$$ \noindent and 
$$\partial_{P_X}(P_Y) = \delta_{XY}$$ 
$$\partial_{P_X}(Y) = 0 = \partial_{X}(P_Y)$$ \noindent with the last equations valid 
even if $X=Y.$ Note also that iterated partial derivatives such as $\partial_X \partial_Y$
commute. Hence the curvature $R_{XY}$ is equal to zero.  We shall regard these position
and momentum 
operators and the corresponding partial derivatives as an abstract algebraic substitute for
flat space.
\bigbreak

With this reference point of (algebraic, quantum) flat space we can define 
$$\hat{P_{X}} = P_{X} - A_{X}$$ 
\noindent for an arbitrary algebra-valued function of the variable
$X.$ In indices this would read
$$\hat{P_{i}} = P_{i} - A_{i},$$ \noindent and with respect to this deformed momentum 
we have the covariant derivative 
$$\nabla_{X}F = [F,\hat{P_Y}] = [F, P_Y + A_Y] = \partial_{Y}F + [F, A_Y].$$
\noindent The curvature for this covariant derivative is given by the formula
 $$R_{XY}F = [\nabla_{X}, \nabla_{Y}]F =  [[\lambda_{X},\lambda_{Y}],F]$$
\noindent where $\lambda_X = P_X - A_X.$  Hence
$$R_{XY} = [P_X - A_X, P_Y - A_Y] = -[P_X,A_Y] - [A_X,P_Y] + [A_X, A_Y]$$
$$= \partial_{X}A_Y - \partial_{Y}A_X + [A_X, A_Y].$$
\noindent With indices this reads
$$R_{ij} = \partial_{i}A_j - \partial_{j}A_i + [A_i, A_j].$$
\noindent and the reader will note that this has the abstract form of the curvature of a
Yang-Mills gauge field, and specifically the form of the electromagnetic field when the potentials
$A_i$ and $A_j$ commute with one another.
\bigbreak

Continuing with this example, we compute
$$[X, \hat{P_Y}] = [X, P_Y - A_Y] = \delta_{XY} - [X, A_{Y}].$$

\noindent Let $$g_{XY} = \delta_{XY} - [X, A_{Y}]$$ \noindent
so that $$[X, \hat{P_Y}] = g_{XY}.$$ We will shortly consider the form of this general case, but
first it is useful to restrict to the case where $[X, A_Y]=0$ so that $g_{XY}=\delta_{XY}.$
This is the domain to which the original Feynman-Dyson derivation applies. In order to enter
this domain, we set $$\dot{X} = DX = \hat{P_X} = P_X - A_X.$$ 

\noindent We then have

$$[X_i,X_j]=0$$
$$[X_i, \dot{X_j}] = \delta_{ij}$$

\noindent and 

$$R_{ij} = [\dot{X_i}, \dot{X_j}] = \partial_{i}A_j - \partial_{j}A_i + [A_i, A_j].$$

\noindent Note that even under these restrictions we are still looking at the possibility of a
non-abelian gauge field. The pure electromagnetic case is when the commutator of $A_i$ and 
$A_j$ vanishes.  But why do we set $\dot{X} = \hat{P_X}?$ The answer to this is the key to 
the gauge interpretation of electromagnetism, for with this interpretation we find that $\dot{X}$
satisfies the Lorentz force law $\ddot{X} = E + \dot{X} \times B$ where $B$ represents the 
magnetic field and $E$ the electric field (in the case of three space variables $X_i$ with
$i = 1,2,3.)$ To see how this works, suppose that $\ddot{X_i} = E_i +F_{ij}\dot{X_j}$ and 
suppose that $E_i$ and $F_{ij}$ commute with $X_k.$ Then we can compute
$$[X_i,\ddot{X_j}] = [X_i, E_j +F_{jk}\dot{X_k}]$$
$$= F_{jk}[X_i, \dot{X_k}] = F_{jk}\delta_{ik} = F_{ji}.$$

\noindent This implies that $$F_{ij} = [\dot{X_i}, \dot{X_j}] = R_{ij} = \partial_{i}A_j - \partial_{j}A_i + [A_i, A_j]$$

\noindent since $[X_i,\ddot{X_j}] + [\dot{X_i}, \dot{X_j}] = D[X_i, \dot{X_j}] = 0.$
It is then easy to verify that the Lorentz force equation is satisfied with 
$B_k = \epsilon_{ijk}R_{ij}$ and that in the case of $[A_i,A_j]=0$ this leads directly to 
standard electromagnetic theory when the bracket is a Poisson bracket (see the next section for
a discussion of Poisson brackets). When this bracket is not zero but the potentials $A_i$ are
functions only of the $X_j$ we can look at a generalization of gauge theory where the 
non-commutativity comes from internal Lie algebra parameters. This shows that the Feynman-Dyson 
derivation supports certain generalizations of classical electromagnetism, and this
will be the subject of a more expanded version of this paper.
\bigbreak

In regard to this last remark, the reader should note that in our \cite{NonCom, Twist} algebraic and discrete version of the 
Feynman-Dyson derivation it was actually an additional assumption that $B \times B = 0$ where $B \times B$
denotes the (non-commutative) vector cross product of $B$ with itself. 
(Note that $B = (1/2) \dot{X} \times \dot{X}.$) In the original Dyson paper this cross product vanished because
of assumptions about the operators and their Hilbert space representations. With $B \times B$ as an extra
term, the Feynman-Dyson derivation is indeed a non-commutative generalization of electromagnetism and includes
forms of gauge theories among its models.
\bigbreak

Generalizing, we wish to examine the structure of the following special axioms for a bracket.
$$[X, DY] = g_{XY}$$
$$[X,Y]=0$$
$$[Z,g_{XY}]=0$$
$$[g_{XY},g_{ZW}]=0$$

Note that $$Dg_{YZ} =D[Y,DZ] = [DY,DZ] + [Y, D^{2}Z].$$ \noindent and that $D[X,g_{XY}]=0$ 
implies that
$$[g_{XY},DZ] = [Z, Dg_{XY}].$$ \noindent
\bigbreak

Define two types of derivations as follows $$\nabla_{X}(F) = [F,DX]$$ \noindent and 
$$\nabla_{DX}(F) = [X,F].$$ \noindent These are dual with respect to $g_{XY}$ and will
act like partials with respect to these variables in the special case when $g_{XY}$ is a 
Kronecker delta, $\delta_{XY}.$ If the form $g_{XY}$ is invertible, then we can rewrite these
derivations by contracting the inverse of $g$ to obtain standard formal partials.
\bigbreak

$$\nabla_{DX}\nabla_{DY} D^{2}Z = [X,[Y,D^{2}Z]]$$
$$= [X,Dg_{YZ} - [DY,DZ]] = [X,Dg_{YZ}] - [X,[DY,DZ]] $$
$$=[g_{YZ},DX] - [X,[DY,DZ]]$$
$$=\nabla_{X}(g_{YZ}) -[X,[DY,DZ]].$$
\noindent Now use the Jacobi identity on the second term and obtain
$$\nabla_{DX}\nabla_{DY} D^{2}Z = \nabla_{X}(g_{YZ}) + [DZ,[X,DY]] + [DY,[DZ,X]] $$
$$= \nabla_{X}(g_{YZ}) - \nabla_{Z}(g_{XY}) + \nabla_{Y}(g_{XZ}).$$ 
\noindent This is the formal Levi-Civita connection.
\bigbreak

At this stage we face once again the mystery of the appearance of the Levi-Civita connection.
There is a way to see that the appearance of this connection is not an accident, but
rather quite natural. We shall explain this point of view in the next section where we discuss
Poisson brackets and the connection of this formalism with classical physics. On the other hand,
we have seen in this section that it is quite natural for curvature in the form of the 
non-commutativity of derivations to appear at the outset in a non-commutative formalism. We have
also see that this curvature and connection can be understood as a measurement of the deviation 
of the theory from the ``flat" commutation relations of ordinary quantum mechanics. 
Electromagnetism and Yang-Mills theory can be seen as the theory of the curvature introduced by 
such a deviation. On the other hand, from the point of view of metric differential geometry, the
Levi-Civita connection is the unique connection that preserves the inner product defined by the 
metric under the parallel translation defined by the connection. We would like to see that the
formal Levi-Civita connection produced here has this property as well.
\bigbreak

To this end lets recall the formalism of parallel translation. The infinitesimal parallel
translate of $A$ is denoted by $A' = A + \delta A$ where 
$$\delta A^{k} = -\Gamma^{k}_{ij}A^{i}dX^{j}$$ \noindent where here we are writing in the usual
language of vectors and differentials with the Einstein summation convention for repeated
indices. We assume that the Christoffel symbols satisfy the symmetry condition
$\Gamma^{k}_{ij} = \Gamma^{k}_{ji}.$ The inner product is given by the formula
$$<A,B> = g_{ij}A^{i}B^{j}$$ \noindent Note that here the bare symbols denote vectors whose 
coordinates may be indicated by indices. The requirement that this inner product be invariant
under parallel displacement is the requirement that $\delta(g_{ij}A^{i}A^{j}) = 0.$
Calculating, one finds 
$$\delta(g_{ij}A^{i}A^{j}) = (\partial_{k}g_{ij})A^{i}A^{j}dX^{k} + g_{ij}\delta(A^{i})A^{j} 
+ g_{ij}A^{i}\delta(A^{j})$$

$$= (\partial_{k}g_{ij})A^{i}A^{j}dX^{k} - g_{ij}\Gamma^{i}_{rs}A^{r}dX^{s}A^{j} 
- g_{ij}A^{i}\Gamma^{j}_{rs}A^{r}dX^{s}$$

$$= (\partial_{k}g_{ij})A^{i}A^{j}dX^{k} - g_{ij}\Gamma^{i}_{rs}A^{r}A^{j}dX^{s} 
- g_{ij}\Gamma^{j}_{rs}A^{i}A^{r}dX^{s}$$

$$= (\partial_{k}g_{ij})A^{i}A^{j}dX^{k} - g_{sj}\Gamma^{s}_{ik}A^{i}A^{j}dX^{k} 
- g_{is}\Gamma^{s}_{jk}A^{i}A^{j}dX^{k}$$

\noindent Hence

$$(\partial_{k}g_{ij}) = g_{sj}\Gamma^{s}_{ik} + g_{is}\Gamma^{s}_{jk}.$$

\noindent From this it follows that 

$$\Gamma_{ijk} = g_{is}\Gamma^{s}_{jk} = (1/2)(\partial_{k}g_{ij} - \partial_{i}g_{jk} 
+\partial_{j}(g_{ik})).$$

\noindent Certainly these notions of variation can be imported into our abstract context.
The question remains how to interpret the new connection that arises. We now have a new 
covariant derivative in the form 

$$\hat{\nabla_{i}}X^{j} = \partial_{i}X^{j} + \Gamma^{j}_{ki}X^{k}.$$

\noindent The question is how the curvature of this connection interfaces with the gauge 
potentials that gave rise to the metric in the first place. The theme of this investigation has
the flavor of gravity theories with a qauge theoretic background. We will investigate these 
relationships in detail in a sequel to this paper.
\bigbreak

\section{Poisson Brackets and Commutator Brackets} 

Dirac \cite{Dirac} introduced a fundamental relationship between quantum
mechanics and classical mechanics that is summarized by the maxim {\em
replace Poisson brackets by commutator brackets.} Recall that the Poisson
bracket $\{ A, B\}$ is defined by the formula 

$$\{ A, B \} = (\partial A/ \partial q) (\partial B/ \partial p) -
(\partial A/ \partial p) (\partial B/ \partial q),$$ 

\noindent where $q$ and $p$ denote classical position and momentum
variables respectively.
\vspace{3mm}

In our version of discrete physics the noncommuting variables are
functions of discrete time, with a $DOC$ derivative $D$ as described in
the first section. Since $DX=XJ -JX = [X,J]$ is itself a commutator, it
follows that $$D([A,B]) = [DA,B] + [A, DB]$$ for any expressions $A$, $B$
in our ring $R$. A corresponding Leibniz rule for Poisson brackets would
read
$$(d/dt) \{ A, B \} = \{ dA/dt , B \} + \{ A, dB/dt \}.$$ 
\vspace{3mm}

\noindent However, here there is an easily verified exact formula:
$$(d/dt) \{ A, B \} = \{ dA/dt , B \} + \{ A, dB/dt \} - \{ A, B
\}(\partial \dot{q} / \partial q + \partial \dot{p} / \partial p).$$ 

\noindent This means that the Leibniz formula will hold for the Poisson
bracket exactly when
$$(\partial \dot{q} / \partial q + \partial \dot{p} / \partial p)=0.$$ 
\vspace{3mm}

\noindent This is an integrability condition that will be satisfied if
$p$ and $q$ satisfy Hamilton's equations
$$ \dot{q} = \partial H / \partial p ,$$ $$ \dot{p} = - \partial H /
\partial q. $$ 
\vspace{3mm}

\noindent This, of course, means that $q$ and $p$ are following a
principle of least action with respect to the Hamiltonian $H$. Thus we
can interpret the {\em fact} $D([A,B]) = [DA,B] + [A, DB]$ in the
discrete (commutator) context as an analog of the principle of least action. Taking
the discrete context as fundamental, we say that Hamilton's equations are
{\em motivated} by the presence of the Leibniz rule for the discrete
derivative of a commutator. The classical laws are obtained by following
Dirac's maxim in the opposite direction! Classical physics is produced by
following the correspondence principle upwards from the discrete.
\vspace{3mm} 

Taking the  last paragraph seriously, we must reevaluate the meaning of
Dirac's maxim.  The meaning of quantization has long been a  basic
mystery of quantum mechanics. By traversing this territory in reverse,
starting from the non-commutative world, we begin these questions anew.
\vspace{3mm} 

In making this backwards journey to classical physics we see how our earlier assertion
that bare quantum mechanics of commutators can be regarded as the background for the
coupling with other fields (as in the description of formal gauge theory in the last section),
fits with Poisson brackets. The bare Poisson brackets satisfy
$$\{q_{i}, q_{j}\}=0$$
$$\{p_{i}, p_{j}\}=0$$
$$\{q_{i}, p_{j}\} = \delta_{ij}.$$ In our previous formalism, we would identify
$X_{i}$ as the correspondent with $q_{i}$  and $P_j$ as the
correspondent of $p_{j}.$ And, given a classical vector potential $A$, we could write the 
coupling $dq_{i}/dt = p_{i} - A_i$ to describe the motion of a particle in the presence of an 
electromagnetic field. The analog of the Feynman Dyson derivation is then expressed classically in 
terms of the Poisson brackets. Similar remarks apply to the analogs for gauge theory and curvature.
In particular it is of interest to see that our derivation of the Levi-Civita connection 
corresponds to the motion of a particle in generalized coordinates that satisfies Hamilton's equations.
The fact that such a particle moves in a geodesic according to the Levi-Civita connection is a classical
fact that was surely one of the motivations for the development of differential geometry.
Our derivation of the Levi-Civita connection, interpreted in Poisson brackets, reproduces this result.
\bigbreak

To see how this works, let $ds^2 = g^{ij}dx_i dx_j$ denote the metric in the generalized coordinates
$x_k.$ Then the velocity of the particle has square $v^2 = (ds/dt)^2 = g^{ij}\dot{x_i}\dot{x_j}.$
The Lagrangian for the system is the kinetic energy $L = mv^{2}/2 = mg^{ij}\dot{x_i}\dot{x_j}/2.$
Then the canonical
momentum is $p_j = \partial L/ \partial\dot{x_j},$ and with $q_i = x_i$ we have the Poisson brakets
$$\delta_{ij} = \{q_i, p_j\} = \{x_i, \partial L/ \partial\dot{x_j}\} = \{x_i, mg^{jk}\dot{x_k}\}.$$ Taking $m=1$
for simplicity, we can rewrite this bracket as $$\{x_i,\dot{x_j}\} = g_{ij}.$$ This, in Poisson brackets, is
our generalized equation of motion. \bigbreak

The classical derivation applies Lagrange's equation of motion to the system. Lagrange's equation reads
$$d/dt(\partial L/\partial \dot{x_i}) = \partial L/\partial x_i.$$ Since this equation is equivalent to 
Hamilton's equation of motion, it follows that the Poisson brackets satisfy the Leibniz rule. With this, we
can proceed with our derivation of the Levi-Civita connection in relation to the acceleration of the particle.
In the classical derivation, one writes out the Lagrange equation and solves for the acceleration.
The advantage of using only the Poisson brackets is that it shows the relationship of the connection with the
Jacobi identity and the Leibniz rule. \bigbreak

This discussion raises further questions about the
nature of the generalization that we have made. Originally Hermann Weyl
\cite{Weyl}  generalized classical differential geometry and discovered
gauge theory by allowing changes of length as well as changes of angle to
appear in the holonomy. Here we arrive at a very similar situation via
the properties of a non-commutative discrete calculus of observations. A
closer comparison with the geometry of gauge theories is called for.
\vspace{3mm}

\section{Discussion on $q$-Deformation} The direct relation between the
content of local physical descriptions based on the $DOC$ calculus and
more global considerations are a matter of speculation.   One strong hint
is contained in the properties of the discrete derivative that has the
form $$D_{q}f(x) = (f(qx) - f(x))/(qx-x).$$ The classical derivative
occurs in the limit as $q$ approaches one. 
\vspace{3mm}

In the setting of $q$ not equal to one, the derivative $D_{q}$ is
directly related to fundamental noncommutativity.   Consider variables
$x$ and $y$ such that $yx=qxy$ where $q$ is a commuting scalar. Then the
expansion of $(x+y)^n$ generates a $q$-binomial theorem with $q$-choice
coefficients composed in $q$-factorials of $q$-integers $[n]_{q}$ where
$$[n]_{q} = 1 + q + q^2 + ... + q^{(n-1)}.$$ The derivative $D_{q}$ is
directly related to the $q$-integers via the formula $$D_{q}(x^{n}) =
[n]_{q} x^{n-1}.$$
\vspace{3mm}

In the context of this paper, we have considered discrete derivatives in
the form $$d_{\Delta}f(x) = (f(x+\Delta) - f(x))/\Delta.$$ This will
convert to the $q$-derivative if $x+ \Delta = qx$.  Thus we need
$$q = (x+ \Delta)/x.$$  This means that a direct translation from $DOC$
to $q$-derivations could be effected if we allowed $q$ to vary as a
function of $x$ and introduced the temporal operator $J$  into the
calculus of $q$-derivatives.  
\vspace{3mm}

In general, many $q$-deformed structures such as the quantum groups
associated with the classical Lie algebras appear to be entwined with the
discretization inherent in $D_{q}.$ The quantum groups have turned out to
be deeply connected with topological amplitudes for networks describing
knots and three dimensional spaces. (See the next section of this paper.)
The analog for the quantum groups in dimension four is being sought. If
there is a connection between the local and the global parts of our essay
it may lie in hidden connections between discretization and quantum
groups. Clearly there is much work to be done in this field.
\vspace{3mm}

There is a clue about the meaning of the operator $J$ ($DF = [F,J]$ in
the discrete ordered calculus)  in the context of quantum groups. 
Quantum groups are Hopf algebras. A quantum group such as $G=U_q(SU(2))$
is actually an algebra over a field $k$ with an antipode $$S:G
\longrightarrow G$$ and a coproduct $$\Delta: G \longrightarrow G \otimes
G,$$ a unit $1$ and a couinit $$\epsilon: G \longrightarrow k.$$ The
coproduct is a map of algebras.  The antipode is an antimorphism,  $S(xy)
= S(y)S(x),$ and generalizes the inverse in a group in the sense that 
$\Sigma S(x_{1})x_{2} = \epsilon(x)1$ and   $\Sigma x_{1}S(x_{2}) =
\epsilon(x)1$ where $\Delta(x) = \Sigma  x_{1} \otimes x_{2}.$ 
\vspace{3mm}

An element $g$ in a quantum group $G$ is said to be a {\em grouplike
element} if $\Delta(g) = g \otimes g$  and $S(g) = g^{-1}.$ In many
quantum groups (such as  $G=U_q(SU(2))$) the square of the antipode is
represented via conjugation by a special grouplike element that we shall
denote by $J$.  Thus  $$S^{2}(x) = J^{-1}xJ$$ for all $x$ in $G.$ This
means that it is possible to define the discrete ordered calculus in the
context of a quantum group $G$ (as above) by taking $J$ to be the special
grouplike element.  Then we have 
$$DX = [X,J] = XJ - JX = J(J^{-1}XJ - X) = J(S^{2}(X) -X).$$ Conjugation
by the special grouplike element in the quantum group constitutes the
time evolution operator in this algebra. 
\vspace{3mm}

There are a number of curious aspects to this use of the discrete ordered
calculus in a quantum group. First of all, it is the case that in some
quantum groups (for example with undeformed classical Lie algebras) the
square of the antipode is equal to the identity mapping. From the point
of view of $DOC$, time does not exist in these algebras. But in the
$q$-deformations such as $U_q(SU(2))$, the square of the antipode is
quite non-trivial and can serve well as the tick of the clock.  In this
way, $q$-deformations do provide a context for time. In particular, this
suggests that the $q$-deformations of classical spin networks
\cite{Pen:Spin} should be able to accommodate time. A suggestion directly
related to this remark occurs in \cite{Crane2}, and we shall take this up
at the end of the next section of this paper.
\vspace{3mm}

\section {Networks, Discrete Spacetime and the Dirac Equation} One can
consider replacing continuous space (such as Euclidean space with the
usual topology) by a discrete structure of relationships. The geometry of
the Greeks held a discrete web of relationships in the context of 
continuous space. That space was not coordinatized in our way, nor was it
held as an infinite aggregate of points. In general topology there is a
wide choice for possible spatial structures (where we mean by a space a
topology on some set). 
\vspace{3mm}

Discretization of space and time implicates the replacement of spacetime
by a network, graph or complex that has nodes for the points and edges to
indicate significant relationships among the points.
\vspace{3mm}

Euler's work in the eighteenth century  brought forth the use of abstract
graphs as holders of spatial structure. After Euler it was possible to
find the classification of the Greek regular solids in the the (wider)
classification of the regular graphs on the surface of the sphere. Metric
can disappear into relationship under the topological constraint of
Euler's formula $V-E+F=2$, where $V$ denotes the number of vertices, $E$
the number of edges and $F$ the number of faces for the connected graph
$G$ on the sphere.
\vspace{3mm}

A network itself can represent an abstract space.  Embeddings of that
network into a given space (such as graphs on the two dimensional sphere)
correspond to global constraints on the structure of the abstract graph. 
\vspace{3mm}

Now a new theme arises, motivated by a conjunction of combinatorics and
physics. Imagine labelling the edges of the network from some set of
``colors". These colors can represent the basic states of a physical
system, or they can be an abstract set of distinct markers for purely
mathematical purposes. Once the network is labelled, each vertex is an
entity with a collection of labels incident to it. Let there be given a
function that associates a number (or algebra element) to each such
labelled vertex. Call this number the {\em vertex weight} at that vertex.
Let $C$ denote a specific coloring of the network $N$ and  consider   the
product, over all the vertices of $N$ of the values of the vertex
weights. Finally let $Z(N)$ , the {\em amplitude} of the network,  be
defined as the summation of  the product of the vertex weights over all
colorings of the net.  $Z(N)$  is  also called the {\em partition
function} of the network. 
\vspace{3mm}

Amplitudes of this sort are exactly what one computes in finding the
partition function of a physical system or the quantum mechanical
amplitude for a discrete process. In all these cases the network is
interwoven with the algebraic structure of the vertex weights. It is only
recently that topological properties of networks in three dimensional
space have come to be understood in this way
\cite {Kauff:KP}, \cite{Atiyah},\cite{Witten:QFTJP}. This has led to new
information about the topology of low dimensional spaces, and new
relationships between physics and topology. 
\vspace{3mm}

A classical example of such an amplitude was discovered by Roger Penrose
\cite {Penrose} in elucidating special colorings of 3-regular graphs in
the plane. A 3-regular graph $G$ has three edges incident to each vertex.
When embedded in the plane, these edges acquire a specific cyclic order.
Three colors are used. One associates to each vertex the weight
$$\sqrt{-1}\ \epsilon_{abc}$$ where $a$,$b$,$c$ denote the edges meeting
the vertex in this cyclic order, and the epsilon is equal to $1$, $-1$
according as the edges have distinct labels in the given or reverse
cyclic order, or $0$ if there is a repetition of labels. The resulting
amplitude counts the number of ways to color the network with three
colors so that three distinct colors are incident to each vertex. This
result is a perspicuous generalization of the classical four color
problem of coloring maps in the plane with four colors so that adjacent
regions receive different colors.
\vspace{3mm}

The Penrose example generalizes to networks whose amplitudes embody
geometrical properties of Euclidean three dimensional space (angles and
their dependence). Geometry begins to emerge in terms of  the  averages
of properties of an abstract and discrete network of relationships.
Topological properties emerge in the same way. The idea of space may 
change to the idea of a network with global states and a functor that
associates this network and its states to the more familiar properties
that a classical observer might see.
\vspace{3mm}

\subsection{Remarks on Quantum Mechanics} 

We should remark on the basic formalism for amplitudes in quantum
mechanics. The Dirac notation $\langle A|B\rangle$ \cite{Dirac} denotes
the probability amplitude for a transition from $A$ to $B$. Here $A$ and
$B$ could be points in space (for the path of a particle), fields (for
quantum field theory), or geometries on spacetime (for quantum gravity).
The probability amplitude is a complex number. The actual probability of
an event is the absolute square of the amplitude. If a complete set of
intermediate states $C_{1}, C_{2},...C_{n}$ is known, then the amplitude
can be expanded to a summation
$$\langle A|B\rangle = \Sigma_{i=1}^{n}\langle A|C_{i}\rangle\langle
C_{i}|B\rangle.$$ This formula follows the formalism of the usual rules
for probability, and it allows for the constructive and destructive
interference of the amplitudes. It is the simplest case of a quantum
network of the form $$A---*---C--- *---B$$ where the colors at $A$ and
$B$ are fixed and we run through all choices of colors for for the middle
edge. The vertex weights at the vertices labelled $*$ are $\langle
A|C\rangle$ and $\langle C|B\rangle$ respectively. A measurement at the
$C$ edge reduces the big summation to a single value.  
\vspace{3mm} 

Consider the generalization of the previous example to the graph

$$A---*---C^{1}---*---C^{2}---*--- ... ---*---C^{m}---B$$

With A and B fixed the amplitude for the net is
$$<A|B> = \Sigma_{1 \leq i_{1} \leq ... \leq i_{m} \leq n}
<A|C^{1}_{i_{1}}><C^{2}_{i_{2}}|C^{3}_{i_{3}}>...<C^{m}_{i_{m}}|B>$$

One can think of this as the sum over all the possible paths from $A$ to
$B.$ In fact in the case of a ``particle" travelling between two points in
space, this is exactly what must be done to compute an amplitude -
integrate over all the paths between the two points with appropriate
weightings.  In the discrete case this sort of summation makes perfect
sense. In the case of a continuum there is no known way to make rigorous
mathematical sense out of all cases of such integrals. Nevertheless, the
principles of quantum mechanics must be held foremost for physical
purposes and so such ``path integrals" and their generalizations to
quantum fields are in constant use by theoretical physicists
\cite{Feynman&Hibbs} who take the point of view that the proof of a
technique is in the consistency of the results with the experiments. 
When the observations themselves are mathematical (such as finding
invariants of knots and links), the issue acquires a new texture.  
\vspace{3mm}

Now consider the summation discussed above in the case where $n=2.$ That
is, we shall assume that each $C^{k} $can take two values, call these
values $L$ and $R.$ Furthermore let us suppose that $<L|R> = <R|L>= \surd
\overline {-1}$ while $<L|L>=<R|R>=1.$ The amplitudes that one computes
in this case correspond to solutions to the Dirac equation \cite{Dirac}
in one space variable and one time variable. This example is related to
an observation of  Richard Feynman  
\cite{Feynman&Hibbs}. In \cite{KN:Dirac} we give a very elementary
derivation of this result and we show how  these amplitudes give
solutions to the discretized Dirac equation, so everything is really
quite exact and one can understand just what happens in taking the limit
to the continuum.  In this example a state of the network consists in a
sequence  of choices of $L$ or $R$. These can be interpreted as choices
to move left or right along the light-cone in a Minkowski plane. It is in
summing over such paths in spacetime that the solution to the Dirac
equation appears.  In this case, time has been introduced into the net by
interpreting the sequence of nodes in the network as a temporal
direction. 
\vspace{3mm}

More specifically, let $(a,b)$ denote a point in discrete Minkowski
spacetime in lightcone coordinates. This means that $a$ denotes the
number of steps taken to the left and $b$ denotes the number of steps
taken to the right. We let $\psi_{L}(a,b)$ denote the sum over the paths
that enter the point $(a,b)$ from the left and $\psi_{R}(a,b)$ the sum
over the paths that enter $(a,b)$ from the right.  Each path $P$ 
contributes $i^{c(P}$ where $c(P)$ denotes the number of corners in the
path. View the diagram below.

{\tt    \setlength{\unitlength}{0.92pt}
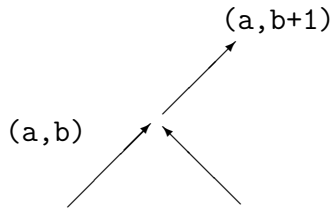
\begin{picture}(158,107)
\thinlines    \put(109,78){\makebox(39,19){(a,b+1)}}
              \put(10,31){\makebox(46,21){(a,b)}}
              \put(113,12){\vector(-1,1){32}}
              \put(81,49){\vector(1,1){30}}
              \put(42,11){\vector(1,1){34}}
\end{picture}}

It is clear from the diagram that
$$\psi_{L}(a, b+1) = \psi_{L}(a, b) + i\psi_{R}(a, b).$$
\noindent Thus we have that
$$\partial \psi_{L} / \partial R = i \psi_{R}$$
\noindent and similarly
$$\partial \psi_{R} / \partial L = i \psi_{L}.$$
\noindent This pair of equations is the Dirac equation in light cone
coordinates.
\vspace{3mm}

This discrete derivation of the Dirac equation is simpler than the method
used in \cite{KN:Dirac}.  I am indebted to Charles Bloom \cite{Bloom} for
pointing this out to me. In fact, this form of the discretization is
essentially Feynman's original method as is evident from the reproduction
of Feynman's handwritten notes in Figure 8 of the review paper
\cite{Schweber} by Schweber.  For one approach, very close in spirit, that generalizes
this exercise of Feynman to four dimensional discrete spacetime see \cite{Smith}.  
\vspace{3mm}

As in the Dirac equation  example, one way to incorporate spacetime is to
introduce a temporal direction into the net. At a vertex, one must
specify labels of {\em before} and {\em after} to each edge of the net
that is incident to that vertex. If there is a sufficiently coherent
assignment of such local times, then a global time direction can emerge
for the entire network. Networks endowed with temporal directions have
the structure of morphisms in a category where each morphism points from
past to future. A category of quantum networks emerges equipped with a
functor (via the algebra of the vertex weights) to morphisms of vector
spaces and representations of generalized symmetry groups. Appropriate
traces of these morphisms produce the amplitudes. 
\vspace{3mm}

Quantum non-locality is built into the network picture. Any observer
taking a measurement in the net has an effect on the global set of states
available for summation and hence affects the possibilities of
observations at all other nodes in the network. By replacing space with a
network we obtain a precursor to spacetime in which quantum mechanics is
built into the initial structure. 
\vspace{3mm} 

\noindent {\bf Remark.} A striking parallel to the views expressed in
this section can be found in \cite{Ett}. Concepts of time and category
are discussed by  Louis Crane \cite{Crane1}, \cite{Crane2} in relation to
topological quantum field theory. In the case of Crane's work there is a
deeper connection with the methods of this paper, as I shall explain
below.
\vspace{3mm}

\subsection{Temporality and the Crane Model for Quantum Gravity}

Crane uses a partition function defined for a triangulated
four-manifold.  Let us denote the partition function by $Z(M^{4}, A, B) =
<A|B>_{M}$ where $M^{4}$ is a four-manifold and $A$ and $B$ are (colored
- see the next sentence) three dimensional submanifolds in the boundary
of $M$. The partition function is constructed by summing over all
colorings of the edges of a dual complex to this triangulation from a
finite set of colors that correspond to certain representations of the
the quantum group $U_{q}(SU(2))$ where $q$ is a root of unity.  The sum
is over products of  $15J_{q}$ symbols (natural generalizations of the
$6J$ symbols in angular momentum theory) evaluated with respect to the
colorings.  The specific form of the partition function (here written in
the case where $A$ and $B$ are empty)  is

$$Z(M^{4}) = N^{v-e} \Sigma_{\lambda} \Pi_{\sigma}
dim_{q}(\lambda(\sigma)) \Pi_{\tau} dim^{-1}_{q}(\lambda(\tau))
\Pi_{\zeta} 15J_{q}(\lambda(\zeta)).$$

Here $\lambda$ denotes the labelling function, assigning colors to the
faces and tetrahedra of $M^{4}$ and $v-e$ is the difference of the number
of vertices and the number of edges in $M^{4}.$  Faces are denoted by
$\sigma$, tetrahedra by $\tau$ and 4-simplices by $\zeta.$ We refer the
reader to \cite{CKY} for further details. 
\vspace{3mm}

In computing $Z(M^{4}, A, B)=<A|B>_{M}$ one fixes the choice of
coloration on the boundary parts $A$ and $B$.  The analog with quantum
gravity is that a colored three manifold $A$ can be regarded as a three
manifold with a choice of (combinatorial) metric. The coloring is the
combinatorial substitute for the metric. In the three manifold case this
is quite specifically so, since the colors can be regarded as affixed to
the edges of the simplices. The color on a given edge is interpreted as 
the generalized distance between the endpoints of the edge. Thus
$<A|B>_{M}$ is a summation over ``all possible metrics" on $M^{4}$ that
can extend the given metrics on $A$ and $B$. $<A|B>_{M}$ is an amplitude
for the metric (coloring) on $A$ to evolve in the spacetime $M^{4}$ to
the metric (coloring) on $B$.
\vspace{3mm}

The partition function  $Z(M^{4}, A, B) = <A|B>_{M}$ is a topological
invariant of the four manifold $M^{4}$. In particular, if $A$ and $B$ are
empty (a vacuum-vacuum amplitude), then  the Crane-Yetter invariant,
$Z(M^{4})$,  is a function of the signature and Euler characteristic of
the four-manifold \cite{CKY}. On the mathematical side of the picture
this is already significant since it provides a new way to express the
signature of a four-manifold in terms of local combinatorial data.
\vspace{3mm}

From the point of view of a theory of quantum gravity, $Z(M^{4}, A, B) =
<A|B>_{M}$, as we have described it so far, is lacking in a notion of
time and dynamical evolution on the four manifold $M^{4}$. One can think
of $A$ and $B$ as manifolds at the initial and final times, but we have
not yet described a notion of time within $M^{4}$ itself.
\vspace{3mm}

Crane proposes to introduce time into $M^{4}$ and into the partition
function  $<A|B>_{M}$ by labelling certain three dimensional submanifolds
of $M^{4}$ with special grouplike elements from the quantum group
$U_{q}(SU(2))$ and extending the partition function to include this
labelling.  Movement across such a labelled hypersurface is regarded as
one tick of the clock.  The special grouplike elements act on the
representations in such a way that the partition function can be extended
to include the extra labels.  Then one has the project to understand the
new partition function and its relationship with discrete dynamics for
this model of quantum gravity.
\vspace{3mm}

Lets denote the special grouplike element in the Hopf algebra  $G =
U_{q}(SU(2))$  by the symbol $J.$  Then, as discussed at the end of the
previous section,  one has that the square of the antipode $S:G
\longrightarrow G$ is given by the formula $S^{2}(x) = J^{-1}xJ.$ This is
the tick of the clock.  The $DOC$ derivative in the quantum group is
given by the formula $DX = [X,J] = J(S^{2}(X) - X).$  I propose to
generalize the discrete ordered calculus on the quantum group to a
discrete ordered calculus on the four manifold $M^{4}$ with its
hyperthreespaces labelled with special grouplikes. This generalised
calculus will be a useful tool in elucidating the dynamics of Crane's
model. Much more work needs to be done in this domain.
\vspace{3mm}

\section{Appendix on Iterants}
The primitive idea behind an iterant is a periodic time series or 
``waveform" $$\cdots abababababab \cdots .$$ The elements of the waveform
can be any mathematically or empirically well-defined objects. We can regard
the ordered pairs $[a,b]$ and $[b,a]$ as abbreviations for the waveform or as two points
of view about the waveform ($a$ first or $b$ first). Call $[a,b]$ an {\em iterant}. 
One has the collection of
transformations of the form $T[a,b] = [ka, k^{-1}b]$ leaving the product $ab$ invariant.
This tiny model contains the seeds of special relativity, and the iterants contain the seeds of 
general matrix algebra! Since this paper has been a combination of discussions of non-commutativity and 
time series, we include this appendix on iterants. A more complete discussion will appear elsewhere.
For related discussion see \cite{SS, SRF, SRCD, IML, KL, BL, Para, GSB}.
\bigbreak 

 Define products and sums of iterants as follows
 $$[a,b][c,d] = [ac,bd]$$  and $$[a,b] + [c,d] = [a+c,b+d].$$
 The operation of juxtapostion is multiplication
 while $+$ denotes ordinary addition in a category appropriate to these entities. These operations are natural 
with respect to the 
 structural juxtaposition of iterants:
 $$...abababababab...$$
 $$...cdcdcdcdcdcd...$$ Structures combine at the points where they correspond.  
Waveforms combine at the times where they correspond. Iterants conmbine in juxtaposition.
\bigbreak

 If $\bullet$ denotes any form of binary compositon for the ingredients 
($a$,$b$,...) of
 iterants, then we can extend $\bullet$ to the iterants themselves by the 
definition
 $[a,b]\bullet[c,d] = [a\bullet c,b\bullet d]$.  In this section we shall first apply this 
idea to Lorentz
 transformations, and then generalize it to other contexts.  
 \vspace{3mm}
 
 So, to work: We have $$[t-x,t+x] = [t,t] + [-x,x] = t[1,1] + x[-1,1].$$  
 Since $[1,1][a,b] = [1a,1b] = [a,b]$  and $[0,0][a,b]= [0,0]$, we shall 
write
 $$1=[1,1]$$ and $$0=[0,0].$$ Let  $$\sigma = [-1,1].$$  $\sigma$ is a 
significant 
 iterant that we shall refer to as a {\em polarity}.  Note that $$\sigma 
\sigma = 1.$$
 Note also that $$[t-x,t+x] = t + x\sigma.$$ Thus the points of 
spacetime form an
 algebra analogous to the complex numbers whose elements are of 
the form $t+x\sigma$
 with $\sigma \sigma = 1$  so that 
 $$(t+x\sigma)(t'+x'\sigma) = tt'+xx' +(tx'+xt')\sigma.$$
 In the case of the Lorentz transformation it is easy to see the 
elements of the form
 $[k,k^{-1}]$ translate into elements of the form

 $$T(v) = [(1+v)/\sqrt{(1-v^{2})}, (1-v)/\sqrt{(1-v^{2})}] = [k,k^{-1}].$$
 
 Further analysis shows that $v$ is the relative velocity of the two 
reference frames in
 the physical context. Multiplication now yields the usual form of the 
Lorentz transform
 $$T_{k}(t + x\sigma) = T(v)(t+x\sigma)$$ 
 $$= (1/\sqrt{(1-v^{2})} -v\sigma/\sqrt{(1-v^{2})})(t+x\sigma)$$ 
 $$=(t-xv)/\sqrt{(1-v^{2})} + (x-vt)\sigma/\sqrt{(1-v^{2})}$$ 
 $$= t' + x'\sigma.$$
 \vspace{3mm}

 The algebra that underlies this iterant presentation of special 
relativity is a relative  
 of the complex numbers with a special element $\sigma$ of square 
one rather than minus
 one ($i^{2} = -1$  in the complex numbers).  
 \vspace{3mm}
 
 The appearance of a square root of minus
 one unfolds naturally from iterant considerations.  Define the ``shift" operator 
$D$ on iterants by the equation $$D[a,b] = [b,a].$$  Sometimes it is 
convenient to think of
 $D$ as a delay opeator, since it shifts the waveform $...ababab...$  
by one internal
 time step. Now define  $$i[a,b] = \sigma D[a,b] = [-1,1][b,a] = [-b,a].$$
 We see at once that $$ii[a,b] = [-a,-b] = [-1,-1][a,b] = (-1)[a,b].$$  
 Thus  $$ii=-1.$$ This is the traditional construction of the square 
root of 
 minus one in terms of operations on ordered pairs.    Here we have 
described $i[a,b]$  in
 a {\em new} way as the superposition of the waveforms $\sigma = [-
1,1]$  and $D[a,b]$
 where  
 $D[a,b]$ is the delay shift of the waveform $[a,b]$.  
\bigbreak

 \subsection{MATRIX ALGEBRA VIA ITERANTS}
Matrix algebra has some strange wisdom built into its very bones.
Consider a two dimensional periodic pattern or ``waveform."
$$......................$$
 $$...abababababababab...$$
 $$...cdcdcdcdcdcdcdcd...$$
 $$...abababababababab...$$
 $$...cdcdcdcdcdcdcdcd...$$
 $$...abababababababab...$$
 $$......................$$
 
$$\left(\begin{array}{cc}
a&b\\
c&d
\end{array}\right), \left(\begin{array}{cc}
b&a\\
d&c
\end{array}\right), \left(\begin{array}{cc}
c&d\\
a&b
\end{array}\right), \left(\begin{array}{cc}
d&c\\
b&a
\end{array}\right)$$ Above are some of the matrices apparent in this array.
\noindent Compare the matrix with the ``two dimensional waveform" shown 
above. A given matrix freezes out a way to view the infinite waveform.
In order to keep track of this patterning, lets write
 
 $$[a,d] + [b,c]\eta = 	\left(\begin{array}{cc}
			a&b\\
			c&d
			\end{array}\right).$$ 

\noindent where
$$[x,y] = 	\left(\begin{array}{cc}
			x&0\\
			0&y
			\end{array}\right).$$ 

\noindent and
$$\eta = 	\left(\begin{array}{cc}
			0&1\\
			1&0
			\end{array}\right).$$

\noindent The four matrices that can be framed in the two-dimensional wave 
form are all obtained from the two iterants
 $[a,d]$ and $[b,c]$ via the delay shift operation $D[x,y] = [y,x]$ which we 
shall denote by  an overbar as shown below  $$D[x,y] = \overline{[x,y]} = [y,x].$$
  
\noindent Letting  $A = [a,d]$  and $B=[b,c]$, we see that the four matrices seen in the 
grid are $$A + B \eta, B + A \eta, \overline{B} + \overline{A}\eta,
  \overline{A} + \overline{B}\eta.$$

 \noindent The operator  $\eta$  has the effect of rotating an iterant by ninety 
 degrees in the formal plane. Ordinary matrix multiplication can be written in a
 concise form using the following rules:

 $$\eta \eta = 1$$
 $$\eta Q = \overline{Q} \eta$$  where  Q is any two element iterant.
 
 \noindent For example, let $\epsilon = [-1,1]$ so that 
$\overline{\epsilon} = - \epsilon$
 and $\epsilon \epsilon = [1,1] = 1.$  
 Let $$i = \epsilon \eta.$$
 
\noindent Then $$ii = \epsilon \eta \epsilon \eta =\epsilon \overline{\epsilon} 
\eta \eta = \epsilon (-\epsilon) = - \epsilon \epsilon = -1.$$
 
 \noindent We have reconstructed the square root of minus one in the form of 
the matrix

   $$ i = \epsilon \eta = [-1,1]\eta 
   =\left(\begin{array}{cc}
			0&-1\\
			1&0
			\end{array}\right).$$
			
\noindent  More generally, we see that 
 $$(A + B\eta)(C+D\eta) = (AC+B\overline{D}) + (AD + 
B\overline{C})\eta$$

 \noindent writing the $2 \times 2$ matrix algebra as a system of 
 hypercomplex numbers.   Note that 
 $$(A+B\eta)(\overline{A}-B\eta) = A\overline{A} - B\overline{B}$$
 
 \noindent The formula on the right corresponds to the determinant of the 
 matrix. Thus we define the {\em conjugate} of
 $A+B\eta$ by the formula 
$$\overline{A+B\eta} = \overline{A} - B\eta.$$
 
\noindent These patterns generalize to higher dimensional matrix algebra.
 \vspace{3mm}

 It is worth pointing out the first precursor to the quaternions: This 
 precursor is the system  $$\{\pm{1}, \pm{\epsilon}, \pm{\eta}, 
\pm{i}\}.$$
 Here $\epsilon\epsilon = 1 = \eta\eta$ while $i=\epsilon \eta$ so 
that $ii = -1$.  
 The basic operations in this
 algebra are those of epsilon and eta.  Eta is the delay shift operator 
that reverses
 the components of the iterant. Epsilon negates one of the 
components, and leaves the
 order unchanged. The quaternions arise directly from these two 
operations once
 we construct an extra
 square root of minus one that commutes with them. Call this extra
 root of minus one $\sqrt{-1}$. Then the quaternions are generated 
by 
 $$\{i=\epsilon \eta, j= \sqrt{-1}\overline{\epsilon}, k=\sqrt{-
1}\eta\}$$  with
 $$i^{2} = j^{2}=k^{2}=ijk=-1.$$
 The ``right" way to generate the quaternions is to start at the bottom 
iterant level
 with boolean values of 0 and 1 and the operation EXOR (exclusive or). Build 
iterants on this,
 and matrix algebra from these iterants.  This gives the square root 
of negation. Now
 take pairs of values from this new algebra and build $2 \times 2$ matrices 
again.  
 The coefficients include square roots of negation that commute with 
constructions at the
 next level and so quaternions appear in the third level of this 
hierarchy. 
\bigbreak

\subsection{Matrix Algebra in General}
Construction of matrix algebra in general proceeds as follows.
Let $M$ be an $n \times n$ matrix over a ring $R.$ Let $M=(m_{ij})$ denote the 
matrix entries. Let $\pi$ be an element of the symmetric group $S_{n}$ so that 
$\pi_1, \pi_2, \cdots , \pi_n$ is a permuation of $1,2,\cdots, n.$ Let 
$v = (v_1, v_2, \cdots, v_n)$ denote a vector with these components.
Let $\Delta(v)$ denote the diagonal matrix whose $i-th$ diagonal entry is $v_i.$
Let $v^{\pi} = (v_{\pi_1},\cdots, v_{\pi_n}).$ Let $\Delta^{\pi}(v) = \Delta(v^{\pi}).$
Let $\Delta$ denote any diagonal matrix and $\Delta^{\pi}$ denote the corresponding permuted
diagonal matrix as just described. Let $[\pi]$ denote the permutation
matrix obtained by taking the $i-th$ row of $[\pi]$ to be the
$\pi_{i} -th$ row of the identity matrix. Note that $[\pi]\Delta = \Delta^{\pi}[\pi].$
For each element $\pi$ of $S_{n}$ define the vector $v(M,\pi) = (m_{1\pi_1},\cdots,m_{n\pi_n})$
and the diagonal matrix $\Delta[M]_{\pi} = \Delta(v(M, \pi)).$
\bigbreak

\noindent {\bf Theorem.} $M = (1/(n-1)!)\Sigma_{\pi \in S_n} \Delta[M]_{\pi} [\pi].$
\bigbreak

\noindent The proof of this theorem is omitted here. Note that the theorem expresses any square 
matrix as a sum of products of diagonal matrices and permutation matrices. Diagonal matrices add and
multiply by adding and multiplying their corresponding entries. They are acted upon by permutations as 
described above. This means that any matrix algebra can be 
embedded in an algebra that has the structure of a group ring of the permutation group with coefficients
$\Delta$ in an
algebra (here the diagonal matrices) that are acted upon by the permutation group, and following the rule
$[\pi]\Delta = \Delta^{\pi}[\pi].$ This is a full generalization of the case $n=2$ described in the last section.
\bigbreak

It is amusing to note that this theorem tells us that up to the factor of $1/(n-1)!$ a unitary matrix
that has unit complex numbers as its entries is a sum of simpler unitary transformations factored into
diagonal and permutation matrices. In quantum computing parlance, such a unitary matrix is a sum of products of
phase gates and products of swap gates (forming the permutations).
\bigbreak 

A reason for discussing these formulations of matrix algebra in the present context is that one sees that 
matrix algebra is generated by the simple operations of juxtaposed addition and multiplication, and by
the use of permutations as operators. These are unavoidable discrete elements, and so the operations of matrix
algebra can be motivated on the basis of discrete physical ideas and non-commutativity. The richness of  
continuum formulations, infinite matrix algebra, and symmetry grows naturally out of 
finite matrix algebra and hence out of the discrete.
\bigbreak 

\section{Philosophical Appendix}
The purpose of this appendix is to point to a way of thinking about the relationship of mathematics, physics, 
persons, and observations that underlies the approach taken in this paper. We began  constructions motivating 
non-commutativity by considering sequences of actions $\cdots DCBA$ written from right to left so that they could
be applied to an actant $X$ in the order $\cdots DCBAX= \cdots (D(C(B(AX))) \cdots.$ The sequence of events
$A,B,C,D, \cdots$ was conceptualized as a temporal order, with the events themselves happening at levels or
frames of successive ``space". {\em There is no ambient coordinate
space, nor is there any continuum of time.} All that is given is the possibility of structure at any given moment, and 
the possibility of distinguishing structures from one moment to the next. In this light the formula 
$DX = [X,J] = XJ - JX = J(X'-X)$ connotes a symbolic representation of the measurment of a difference across one time
interval, nothing more. In other words $DX$ represents a difference taken across a background difference (the time step).
Once the pandora's box of measuring such differences has been opened, we are subject to the multiplicities of forms of 
difference $\nabla_{K}X = [X,K]$, their non-commutativity among themselves, the notion of a flat background that has
the formal appearance of quantum mechanics, the emergence of abstract curvature and formal gauge fields. All this occurs
in these calculi of differences {\em prior} to the emergence of differential geometry or topology or even the notion of 
linear superposition of states (so important to quantum mechanics). Note that in this algebraic patterning each algebra
element $X$ is an actant (can be acted upon) and an actor (via the operator $\nabla_{X}$). In Lie algebras, this
is the relationship between the algebra and its adjoint representation that makes each element of the algebra into a
representor for that algebra by exactly the formula $adj_{A}(X) = [A,X] = - \nabla_{A}(X)$ that we have identified
as a formal difference or derivative, a generator for a calculus of differences.
\bigbreak

The precursor and conceptual background of our particular formalism is therefore the concept of discrimination,
the idea of a distinction. A key work in relation to that concept is the book ``Laws of Form" by G. Spencer-Brown
\cite{GSB} in which is set out a calculus of distinction of maximal simplicity and generality. In that calculus
a mark (denoted here by a bracket $<~>$) represents a distinction and is seen to be a distinction between  inside
and outside. In this elemental mathematics there is no distinction
except the one that we draw between the mathematician and the operator in the formal system as 
sign/symbol/interpretant. This gives full responsibility to the mathematician to draw the boundaries between the formal
system as physical interaction and the formal system as symbolic entity and the formal system as Platonic conceptual form.
In making a mathematics of distinction, the mathematician tells a story to himself/herself about the creation of
a world. Spencer-Brown's iconic mathematics can be extended to contact any mathematics, and when this happens that 
mathematics is transformed into a personal creation of the mathematician who uses it. In a similar (but to a mathematician)
darker way, the physicist is intimately bound to the physical reality that he studies.
\bigbreak
 
We could have begun this paper 
with the the Spencer-Brown mark as bracket: $<~>.$ This empty bracket is seen to make a distinction
between inside and outside. In order for that to occur the bracket has to
become a process in the perception of someone. It has to leave whatever
objective existence or potentiality it has alone (all one) and become the locus or
nexus of an idea in a perceiving mind. As such it is stabilized by that
perception/creation and becomes really a solution to $\{<~>\} = <~>$ where
the curly bracket (the form of perception) is in the first place identical to the mark $<~>,$ and then
distinguished from it by the act of distinguishing world and perceiver.
It is within this cleft of the infinite recursive and the finite
$$<~> = \{<~>\} = \{\{<~>\}\} = \{\{\{<~>\}\}\} = \cdots =  \{\{\{\{\{\{\cdots \}\}\}\}\}\} $$ that the
objectivity of mathematics/physics (they are not different in the cleft) arises.  All the rest of mathematics or
calculus of brackets needs come forth for the observer in the same way.
Through that interaction there is the possibility of a deep dialogue of
many levels, a dialogue where it is seen that mathematics and physics
develop in parallel, each describing the same boundary from opposite sides.
That boundary is the imaginary boundary between the inner and outer worlds of an individual.


\begin{thebibliography}{99}

\bibitem{Atiyah} M.F. Atiyah [1990], {\em The Geometry and Physics of
Knots,}  Cambridge University Press.

\bibitem{BK}
Bastin, T. and Kilmister C. [1995], {\em Combinatorial Physics}, World Scientific Pub. Co.

\bibitem{Bloom} Bloom, Charles [1998], (private communication)

\bibitem{Dyson}
Dyson, F. J. [1990], Feynman's proof of the Maxwell Equations, {\em Am. J. Phys.} 58 (3), March 1990,
209-211.

\bibitem{Penrose} R. Penrose [1971], Applications of negative dimensional
tensors, In {\em Combinatorial Mathematics and Its Applications}, edited
by D. J. A. Welsh, Academic Press.

\bibitem{CKY} Crane, Louis , Kauffman, Louis H., Yetter, David N. [1997],
State sum invariants of 4-manifolds, {\em Journal of Knot Theory and Its
Ramifications}

\bibitem{Connes} Connes,Alain [1990],  {\em Non-commutative Geometry}
Academic Press.

\bibitem{Crane1} Crane,Louis [1996],  Clock and category: Is quantum
gravity algebraic?, {\em J. Math. Phys.} {\bf 36} (11), November (1996),
pp. 6180-6193. 

\bibitem{Crane2} Crane, Louis [1997],  A proposal for the quantum theory
of gravity, arXiv:gr-qc/9704057 v2 23 Apr 97.

\bibitem{Dimakis}
Dimakis, A. and M\"{u}ller-Hoissen [1992], F., Quantum mechanics on a lattice and q-deformations, 
{\em Phys. Lett.} 295B, p.242.


\bibitem{Dirac} Dirac, P.A.M. [1968],  {\em Principles of Quantum
Mechanics,} Oxford University Press.

\bibitem{Ett} Etter, T. and Noyes, Pierre [2001], Process, System, Causality and Quantum
Mechanics, {\em Bit String Physics}, edited by H. Pierre Noyes and J. C. van den Berg,
World Scientific Pub. Co., 488-537.  

\bibitem{Forgy}
Forgy,Eric A. [2002]  Differential geometry in computational electromagnetics, 
PhD Thesis, UIUC.

\bibitem{Fkin} E. Fredkin [1990], Digital Mechanics, {\em Physica D} {\bf
45}, pp. 254-270. 

\bibitem{Smolin:QG} Ashtekar,Abhay, Rovelli, Carlo and Smolin,Lee [1992],
"Weaving a Classical Geometry with Quantum Threads", {\em Phys. Rev.
Lett.}, vol. 69, p. 237. 


\bibitem{Feynman&Hibbs} R.P. Feynman and A.R. Hibbs [1965], {\em Quantum
Mechanics and Path Integrals,} McGraw Hill Book Company.

\bibitem{Hughes}
Hughes, R. J. [1992], On Feynman's proof of the Maxwell Equations, {\em Am. J. Phys.} 60, (4),
April 1992, 301-306.
 
\bibitem{SS}
Kauffman, L. [1985], Sign and Space,  In Religious Experience and Scientific Paradigms. Proceedings of
the 1982 IASWR Conference, Stony Brook, New York: Institute of Advanced Study of
World Religions, (1985), 118-164.

\bibitem{SRF}
Kauffman, L. [1987], Self-reference and recursive forms,  Journal of Social and Biological Structures 
(1987), 53-72.

\bibitem{SRCD}
Kauffman, L. [1987], Special relativity and a calculus of distinctions.  Proceedings of the 9th Annual
Intl. Meeting of ANPA, Cambridge, England (1987).  Pub. by ANPA West, pp.
290-311.

\bibitem{IML}
Kauffman, L. [1987], Imaginary values in mathematical logic.  Proceedings of the Seventeenth
International Conference on Multiple Valued Logic, May 26-28 (1987), Boston MA,
IEEE Computer Society Press, 282-289.

\bibitem{Kauff:KP} Kauffman,Louis H.[1991,1994], {\em Knots and Physics,}
World Scientific Pub.

\bibitem{KL} 
L. H. Kauffman, Knot Logic,  In {\it Knots and Applications}  ed. by L. Kauffman, World Scientific Pub. Co.,
(1994), 1-110.
 

\bibitem{KN:QEM} Kauffman,Louis H. and Noyes,H. Pierre [1996], Discrete
Physics and the Derivation of Electromagnetism from the formalism of
Quantum Mechanics, {\em Proc. of the Royal Soc. Lond. A}, {\bf 452}, pp.
81-95. 

\bibitem{KN:Dirac} Kauffman,Louis H. and Noyes,H. Pierre [1996], Discrete
Physics and the Dirac Equation, {\em Physics Letters A}, 218 ,pp.
139-146. 

\bibitem{KN:DG} Kauffman,Louis H. and Noyes,H.Pierre (In preparation)

\bibitem{Twist} Kauffman, Louis H.[1996], Quantum electrodynamic
birdtracks, {\em Twistor Newsletter Number 41} 

\bibitem{NonCom} Kauffman, Louis H. [1998], Noncommutativity and discrete
physics, {\em Physica D } 120 (1998), 125-138.

\bibitem{ST} Kauffman, Louis H. [1998], Space and time in discrete physics, 
{\em Intl. J. Gen. Syst.} Vol. 27, Nos. 1-3, 241-273.

\bibitem{Aspects}
Kauffman, Louis H. [1999], A non-commutative approach to discrete physics, 
in {\em Aspects II - Proceedings of ANPA 20}, 215-238.

\bibitem{BL}
Kauffman, Louis H. [2002], Biologic. {\em AMS Contemporary Mathematics Series},
Vol. 304, (2002), pp. 313 - 340.

\bibitem{Para}
Kauffman, Louis H. [2002], Time imaginary value, paradox sign and space,
in {\em Computing Anticipatory Systems, CASYS -
Fifth International Conference}, Liege, Belgium (2001) ed. by Daniel Dubois,
AIP Conference Proceedings Volume 627 (2002).

\bibitem{Manthey} 
Manthey, M. [1999], A combinatorial bit-bang leading to the quaternions, Aspects II - Proceedings
of ANPA 20, ed. by Keith Bowden, 23-45.

\bibitem{Mont}
Montesinos, M. and Perez-Lorenzana, A., [1999], Minimal coupling and Feynman's proof,
arXiv:quant-phy/9810088 v2 17 Sep 1999.


\bibitem{MH}
M\"{u}ller-Hoissen,Folkert [1998],  Introduction to non-commutative geometry of commutative algebras and 
applications in physics, in {\em Proceedings of the 2nd Mexican School on Gravitation and Mathematical
Physics}, Kostanz (1998) <http://kaluza.physik.uni-konstanz.de/2MS/mh/mh.html>.


\bibitem{Noyes} Noyes, Pierre [2001], {\em Bit String Physics}, 
edited by H. Pierre Noyes and J. C. van den Berg,
World Scientific Pub. Co.  

\bibitem{micro}
Noyes, H. P., Kauffman, L.H., Lindsay, J.V., Lamb, W.R. [2003] , On E.D. Jones' microcosmology,
SLAC-PUB-9620, Jan. 2003, arXiv:astro-ph/0301176 v1 10 Jan 2003. 


\bibitem{Pen:Spin} Penrose,Roger [1971], Angular Momentum - An Approach
to Combinatorial Spacetime, In {\em Quantum Theory and Beyond},Edited by
Ted Bastin Cambridge University Press, pp. 151-180. 

\bibitem{Schweber} Schweber, Silvan S. [1986], Feynman and the
visualization of space-time processes, {\em Rev. Mod. Phys.} Vol. 58, No.
2, April 1986, 449 - 508.

\bibitem{Smith}
Smith, T., [1997] From sets to quarks, hep-ph/9708379 17 Aug 1997  via 
 http://www.innerx.net/personal/tsmith/TShome.html

\bibitem{GSB} 
G. Spencer-Brown, {\it Laws of Form,} Julian Press, New York (1969).


\bibitem{Tanimura} Tanimura,Shogo [1992], Relativistic generalization and
extension to the non-Abelian gauge theory of Feynman's proof of the
Maxwell equations, {\em Annals of Physics, vol. 220}, pp. 229-247. 

\bibitem{Wene}    
Wene, G. P. [1982], A little color in abstract algebra,{\em Amer. Math. Monthly}, Vol. 89
No. 6 (June - July 1982), 417-419.

\bibitem{Weyl} Weyl, Hermann [1922], {\em Space--Time--Matter}, Methuen,
London (1922).

\bibitem{Witten:QFTJP} Witten,Edward [1989], Quantum field Theory and the
Jones Polynomial, {\em Commun. Math. Phys.},vol. 121,pp. 351-399. 

\end{thebibliography}
\end{document}